\documentclass[showpacs]{revtex4}
\usepackage{graphicx}
\usepackage{dcolumn}
\usepackage{amsmath}
\usepackage{color}
\usepackage{epstopdf}
\usepackage{graphicx}
\usepackage{subfigure}
\usepackage[utf8x]{inputenc}
\definecolor{coolblack}{rgb}{0.0, 0.18, 0.39}
\definecolor{darkred}{rgb}{0.5,0,0}
\definecolor{darkgreen}{rgb}{0,0.5,0}
\definecolor{darkblue}{rgb}{0,0,0.5}
\definecolor{lapislazuli}{rgb}{0.15, 0.38, 0.61}
\definecolor{venetianred}{rgb}{0.78, 0.03, 0.08}
\definecolor{bleudefrance}{rgb}{0.19, 0.55, 0.91}
\definecolor{dogwoodrose}{rgb}{0.84, 0.09, 0.41}
\usepackage[linktocpage,colorlinks]{hyperref}
\hypersetup{colorlinks=true, citecolor=darkgreen, linkcolor=blue,urlcolor = darkblue}
\makeatletter
\def\btt#1{\texttt{\@backslashchar#1}}
\DeclareRobustCommand\bblash{\btt{\@backslashchar}} \makeatother

\RequirePackage{fix-cm}

\begin{document}
\title{Optical and Thermodynamic Properties of a Rotating Dyonic Black Hole Spacetime in $\mathcal{N} = 2, U(1)^2$ gauged supergravity
}
\author{Prateek Sharma $^{a}$}\email{prteeksh@gmail.com}
\author{ Hemwati Nandan $^{a,b}$}\email{hnandan@associates.iucaa.in}
\author{Uma Papnoi$^{a}$}\email{uma.papnoi@gmail.com}
\author{Arindam Kumar Chatterjee $^{a}$}\email{72arindam@gmail.com}
\affiliation{$^{a}$Department of Physics, Gurukula Kangri Vishwavidyalaya, Haridwar 249 404, Uttarakhand, India}
\affiliation{$^{b}$Center for Space Research, North-West University, Mahikeng 2745, South Africa}


\begin{abstract}
 \noindent The null geodesics and the distance of closest approach for photon around a rotating dyonic black hole in $\mathcal{N} = 2, U(1)^2$ gauged supergravity is studied. The phenomenon of black hole shadows with various black hole parameters has also analyzed.  Further, the investigation of various thermodynamic properties for this black hole is performed with various thermodynamic parameters at the horizon. The heat capacity to study the thermodynamic stability of this black hole spacetime is also studied. The influence for different values of the black hole parameters $ \nu $, $ e $, $ \nu $, $ g $ and $N_{g}$ on the phenomenon of black hole shadows and thermodynamic parameters is also investigated visually. 
\end{abstract}
\maketitle
\section{Introduction}
\noindent The AdS/CFT correspondence which is also known as gauge/gravity duality has been emerged in view of the framework of string theory\cite{de2016conceptual}.  These theories can be proved to be a significant step to understand gravity within the regime of quantum theory. In symmetry theories, the $ \mathcal{N} = 2 $ supergravity theory has more symmetries than $ \mathcal{N} = 1 $, but at the same time less constrained than higher N theories (such as the maximal $ \mathcal{N} = 8 $ supergravity). It is also well known that the  theories with a low number of supersymmetries ( $ \mathcal{N} = 1, 2 $) are closer to the standard model. An interesting spacetime rotating black hole (BH) solution in 4 dimensions  $ \mathcal{N} = 2 $, $ U(1)^{2} $ gauged supergravity which can play a crucial role to study the nature of AdS geometry is founded by Chow and Compère \cite{chow2014dyonic}. This BH spacetime has many sub cases with pairwise equal charges and vanishing Newman-Unti-Tamburino (NUT) charge. The detailed analytical study of the geodesics for this BH spacetime has already been performed by Kai Flathmann and Saskia Grunau in \cite{flathmann2016analytic}. The study of physical aspects of BH in strong gravity regime can be used as a probe of quantum effects of any fundamental theory, such as supergravity theories. It is therefore interesting to study the thermodynamic aspects \cite{MahulikarShripadP.2013,SadeghiJ.2014,MyungYunSoo2008,BelhajA.2016,SureshJishnu2014,MiaoYanGang2016,Parikh2017,Smolin2017,Ali2019,Hennigar2019,Appels2016} and phenomenon of BH shadow of this BH spacetime in the context of supergravity.\\
One of the direct observations of BH is to observe its shadow. The study of phenomenon for BH shadow can provide the useful information about the various physical aspects of any given BH spacetime and the recent image from Event Horizon Telescope (EHT) \cite{collaboration2019first,akiyama2019II,akiyama2019III,akiyama2019iv,akiyama2019v,akiyama2019vi} marks a significant mile stone in the study of BH physics. The mathematical aspect of the very phenomen-on which has been developed over the year \cite{synge1966escape,grenzebach2014photon,papnoi2014shadow,StuchlikZdenek2018,AtamurotovFarruh2016,AmirMuhammed2018,AbdujabbarovAhmadjon2015,HuangYang2016,Amarilla2012,BisnovatyiKogan2018,Moffat2020,Contreras2020,Chang2020,Dey2020,Cunha2015} can be studied with observed data which can further help to hone the mathematical models of the extraordinary astronomical objects like a BH. Based on such standpoints, we are motivated to study the motion of photons in ADS geometry to observe the effect of different charges on the path of the light along with the shadows and the relation for distance of closest approach with the magnetic  charges (v), the gauge coupling constant (g) and NUT charge ($\mathcal{N}_g$).
The present paper has an outline in the  form as below.\\ In section 2, we have briefly discussed the geodesics of the rotating dyonic BH in $\mathcal{N} = 2$, $U(1)^2$ gauged supergravity spacetime discussed above. The shadows of this BH are investigated in section 3. Further, the investigation of various thermodynamic properties for  BH with dyonic charges is also studied. We have obtained various thermodynamic parameters at the horizon. The heat capacity to perform the thermodynamic stability of this BH spacetime is also calculated. The variations of all variables with horizon radius at different values of the $ \nu $, $ g $ and $\mathcal{N}_g$ are visually presented in section 4. The  results obtained has been discussed in section 5. 

\section{Rotating Dyonic BH spacetime in $\mathcal{N} = 2, U(1)^2$ gauged supergravity}\label{hsknads}
The spacetime corresponding to the general solution for $\mathcal{N} = 2, U(1)^2$ gauged supergravity BH 
with dyonic charges \cite{chow2014dyonic,flathmann2016analytic,Rudra:2019ssz} is described by the metric.

\begin{eqnarray}\label{metric}
	ds^2 = -\frac{R_g}{B-a A}(dt - \frac{A}{\Xi}d\phi)^2+\frac{B-a A}{R_g}dr^2+\frac{\Theta_g a^2 \sin^2\theta}{B-a A}(dt- \frac{B}{a \Xi} d\phi)^2+\frac{B-a A}{\Theta_g}d\theta^2, \label{metric}
\end{eqnarray}

	\noindent where, $
	R_g=r^2-2 m r +a^2+e^2-N_g^2+g^2(r^4+(a^2+6 N_g^2- 2 v^2)r^2+3 N_g^2(a^2- N_g^2))
	$, $ A= a \sin^2\theta+ 4 N_g \sin^2(\theta/2)$,
	$
	\Theta_g =1-a^2 g^2 \cos^2\theta- 4 a^2 N_g \cos\theta, \;\;\;\;\;\;	B = r^2 +(N_g+a)^2 -v^2	$ and$ \;\;\;  \Xi = 1-4 N_g a g^2 -a^2 g^2$.

The BH spacetime mentioned above in \eqref{metric}, has in general six hairs which are parameterized by mass (m), rotation (a), electric charges (e), magnetic charges (v), the gauge coupling constant (g) and NUT charge ($\mathcal{N}_{g}$).\\
The Lagrangian appropriate for the metric given by \eqref{metric} at equatorial plane is as follows, 

\begin{widetext}
\begin{equation}
	2 \mathcal{L} = -\left[\frac{R_{g} - a^{2}}{(B - a A)}\right] \dot{t}^{2} +\left[\frac{B^{2}- R_{g} - A^{2}}{\Xi^{2}(B - a A)}\right] \dot{\phi}^{2} -2\left[\frac{a B- A R_{g}}{\Xi (B - a A)}\right] \dot{t} \dot{\phi} + \left[\frac{(B - a A)}{R_{g}}\right] \dot{r}^{2},
\end{equation}
\end{widetext}%

and the first integral may therefore be written accordingly,
\begin{equation}
	\dot{t} = \left[\left(\frac{B^{2}- R_{g} A^{2}}{\Xi^{2}(B - a A)}\right) E - \left(\frac{a B- A R_{g}}{\Xi (B - a A)}\right)L_{z} \right]\frac{\Xi^{2}}{R_{g}},
\end{equation}
\begin{equation}
	\dot{\phi} = \left[\left(\frac{aB - R_{g} A}{\Xi (B - a A)}\right) E - \left(\frac{R_{g} - a^{2}}{(B - a A)}\right)L_{z} \right]\frac{\Xi^{2}}{R_{g}},
\end{equation}
\begin{equation}
	\dot{r} = \sqrt{\dfrac{\left(B E - a \Xi L_{z}\right)^{2}-\left(A E -\Xi L_{z} \right)^{2} R_{g}}{(B - a A)^{2}}}. \label{rdot}
\end{equation}
Here $ E $ and $ L_{z} $  are the energy and the angular momentum of the photon respectively.

\begin{figure*}[ht]
	\begin{center}
		
		\begin{tabular}{|c|c|c|}\hline
			\includegraphics[width= 5.5 cm, height= 4cm]{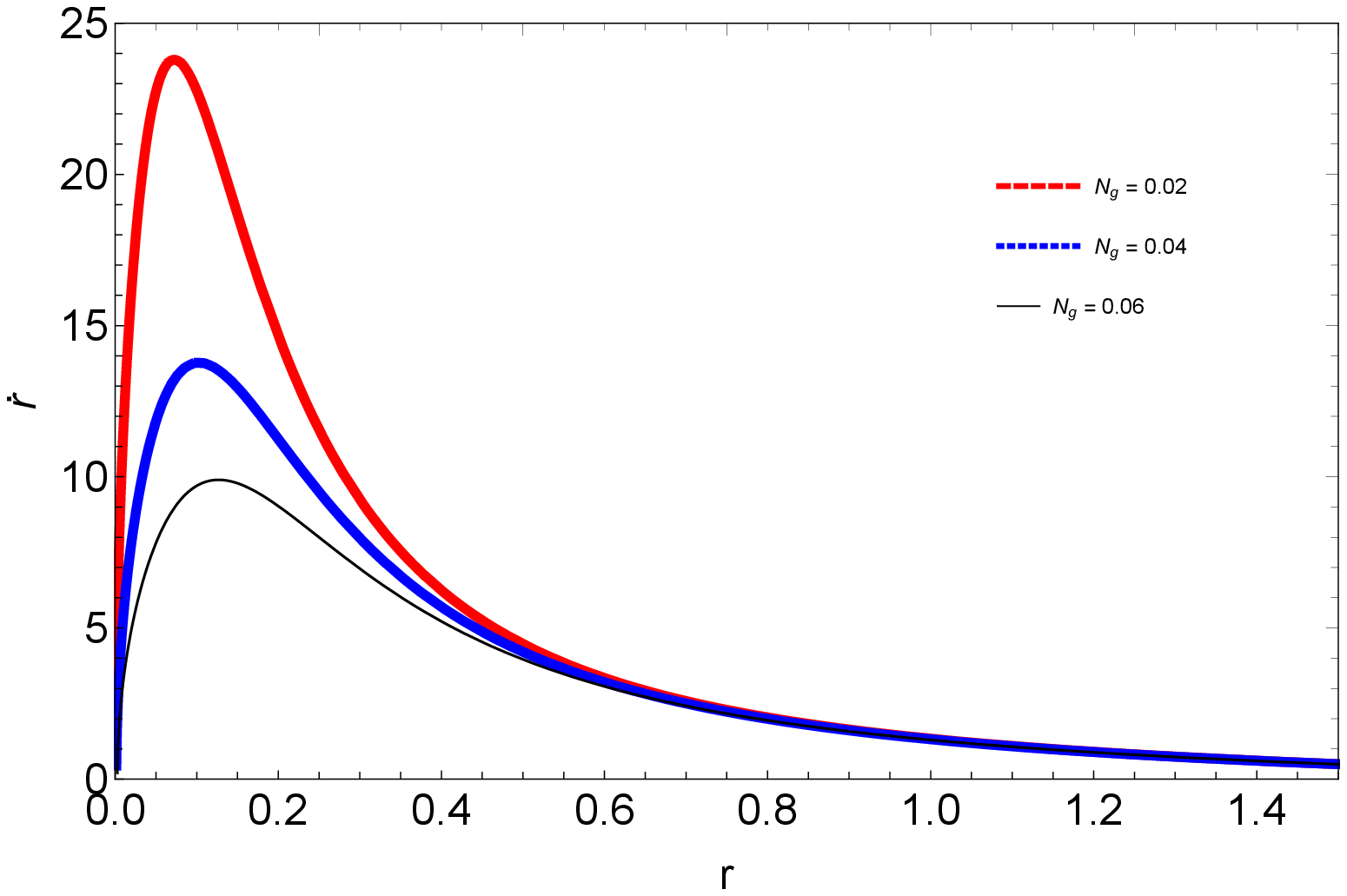}&
			\includegraphics[width= 5.5 cm, height= 4 cm]{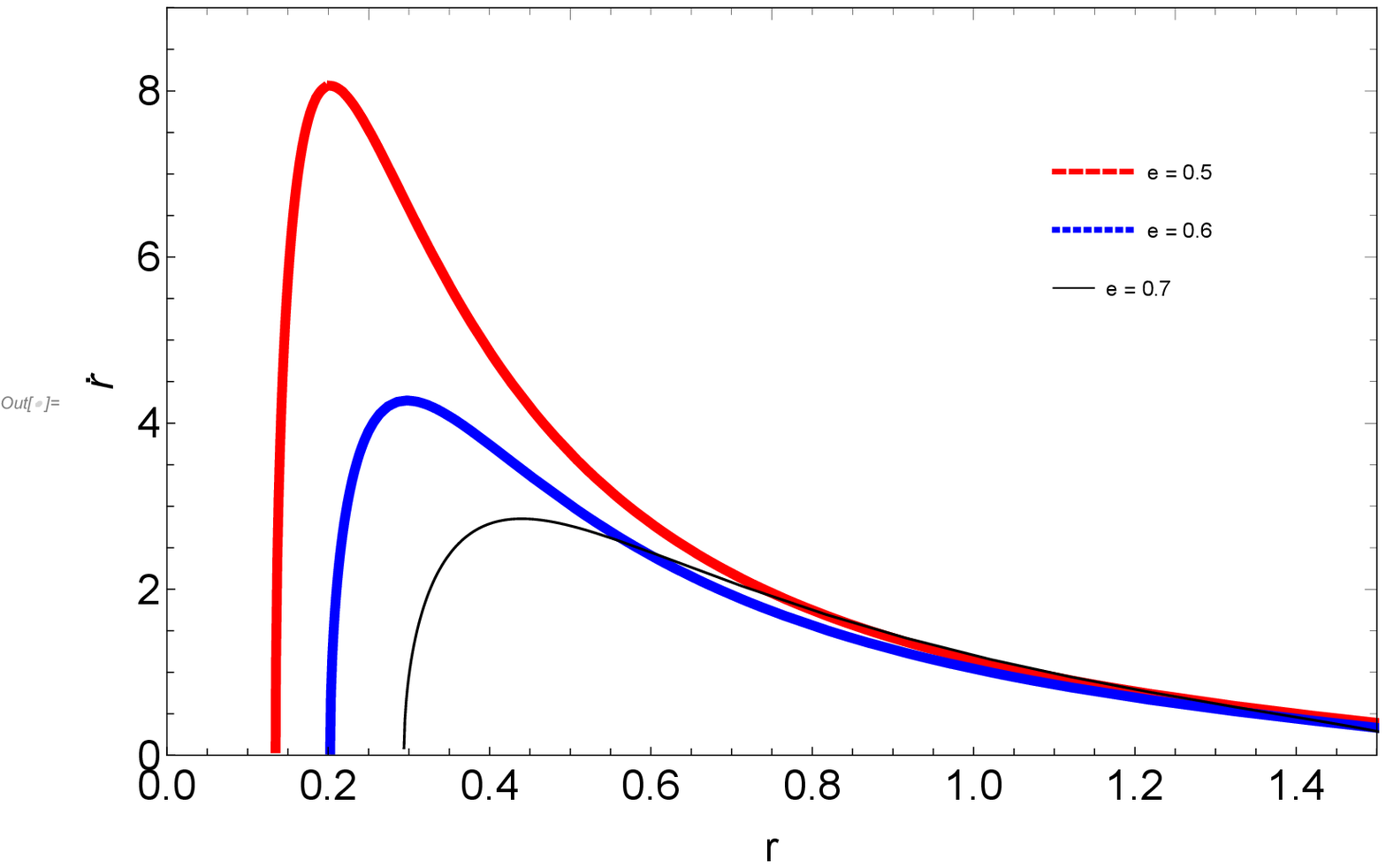}&
			\includegraphics[width= 5.5 cm, height= 4 cm]{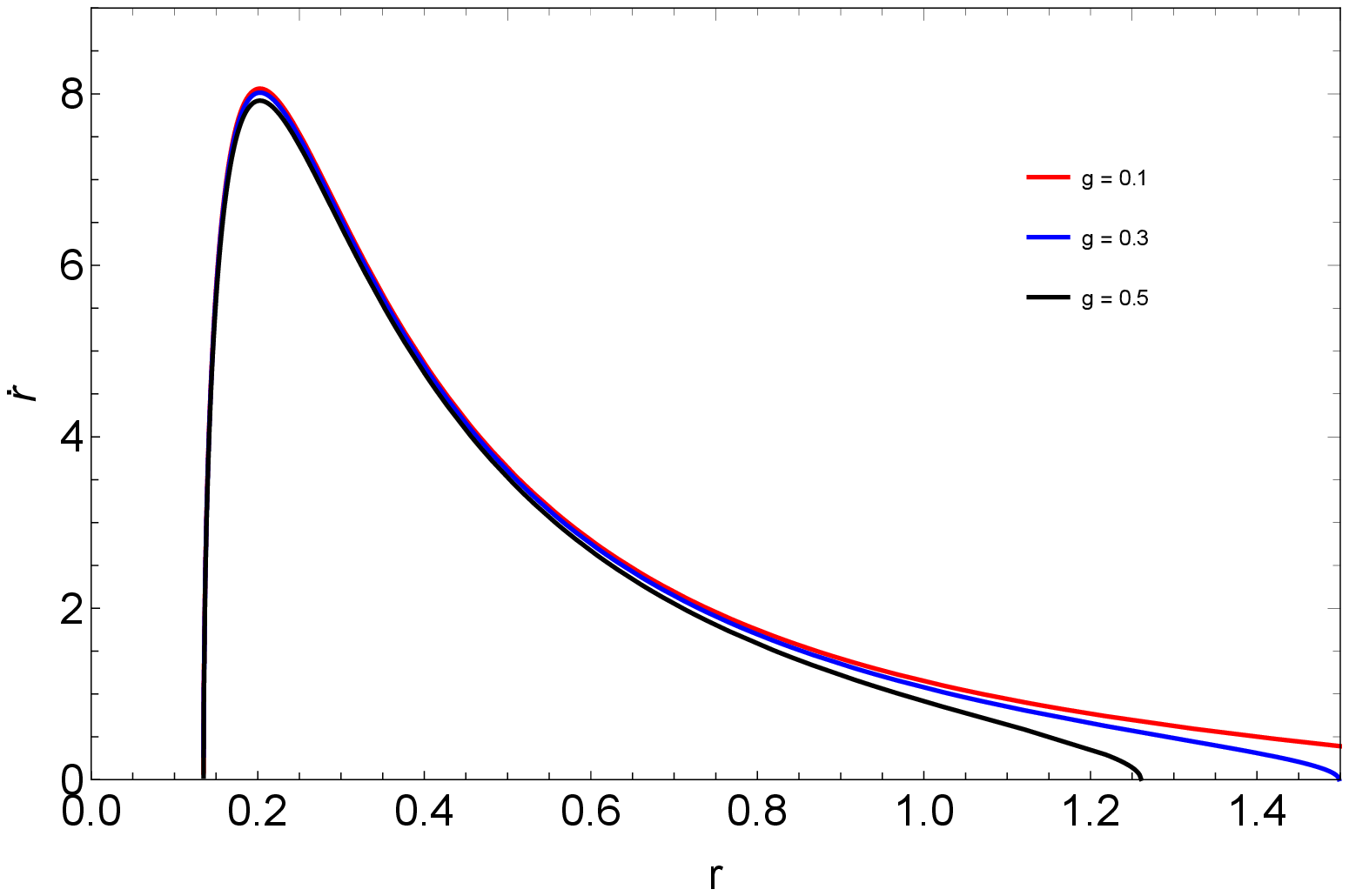}\\
		
			\hline
			
		\end{tabular} 
		\caption{\label{vefff} The variation of $ \dot{r} $ for different values of   NUT charge $\mathcal{N}_{g}$ (left), electric charge $\nu$ (middle), gauge coupling constant $g$  (right), here  $M=1$, $ a = 1, L_{z} = 2.0$ and $E = 0.1 $.}
	\end{center}
\end{figure*}

\noindent From \figurename{\ref{vefff}}, it may be observed that as the value of $ \mathcal{N}_{g} $, $ e $ increase, the peak value of $ \dot{r} $ curve decreases which implies that the maximum value of kinetic energy attained by photons decreases. However, in case of $ g $, the change in peak value is very trivial.   
Further, the condition to have the distance of closest approach is $ \dot{r}  = 0$ and putting the value of $ \dot{r} $ in \eqref{rdot}, one can obtain a quatic equation, which is solved analytically and the possible solutions are given below,
\begin{equation}\label{rp}
	r_{+} = \dfrac{\sqrt{2 \epsilon} \pm \sqrt{-(2 p + 2 \epsilon+\sqrt{\frac{2}{\epsilon}}q)}}{2},
\end{equation}
and
\begin{equation}\label{rn}
	r_{-} = \dfrac{-\sqrt{2 \epsilon} \pm \sqrt{-(2p + 2 \epsilon - \sqrt{\frac{2}{\epsilon}}q)}}{2},
\end{equation}
where, 
\begin{equation*}
	\epsilon = 2 \sqrt{\dfrac{\frac{p^{2}}{12}-s}{3}}\left[\frac{1}{3} \arccos\left(\dfrac{4 \left(\frac{p^{3}}{108}-\frac{p s}{3}-\frac{q^{2}}{8}\right)}{\left(\dfrac{\frac{p^{2}}{12}-s}{3}\right)^{\frac{3}{2}}} \right)\right] - p,
\end{equation*}
\begin{widetext}
\begin{equation*}
	p = \dfrac{\left[2 E\left( \gamma E - a\Xi L_{z} \right)+\left(A E-\Xi L_{z}\right)^{2}\left(1+g^{2}\left(a^{2} + + 6N_{g}^{2}-2\nu^{2} \right)\right)\right]}{\left[E^{2} - \left(A E-\Xi L_{z}\right)g^{2}\right]},
\end{equation*}
\end{widetext}

\begin{equation*}
	q = \dfrac{2 m \left(A E-\Xi L_{z}\right)^{2} }{\left[E^{2} - \left(A E-\Xi L_{z}\right)g^{2}\right]},
\end{equation*}

\begin{equation*}
	s = \dfrac{a^{2} + e^{2} - N_{g}^{2} + 3 g^{2} N_{g}^2 \left(a^{2} - N_{g}^{2}\right) + \left(\gamma E -a \Xi L_{z}\right)^{2}}{E^{2}- \left(A E - \Xi L_{z}\right) g^{2}},
\end{equation*}
and
\begin{equation*}
	\gamma = (N_{g} + a)^{2}-\nu^{2}.
\end{equation*}

The equations  \eqref{rp} and \eqref{rn} represent two roots for the distance of closest approach for the photons. 

\section{Shadow of the Black hole}
In this section, we perform the study of shadow for BH spacetime given by \eqref{metric}, for a locally non-rotating observer. Here  the inclination angle is considered in equatorial plane (i.e. $ \theta  = \frac{\pi}{2} $). The methodology for the calculation here is inclined with the detailed investigation in  \cite{grenzebach2014photon,papnoi2014shadow,bambi2009apparent}. The Hamilton-Jacobi equation governing the motion of photons for the spacetime given in \eqref{metric}  can be describe by,
\begin{equation}\label{key}
	2\frac{\partial S}{\partial \tau} = g^{ij}\frac{\partial S}{\partial x^{i}}\frac{\partial S}{\partial x^{j}},
\end{equation}
here $ S $ is Hamilton's principle function and $ g^{ij} $ is the metric tensor. The equations which describe motion of photons are as below,
\begin{equation}\label{rdot}
	\dot{r} (B-a A_{\pi/2}) = \pm \sqrt{\mathcal{R}} ,
\end{equation}   
\begin{equation}\label{thetadot}
	\dot{\theta} \sin^{2}(\theta) (B-a A_{\pi/2}) = \pm \sqrt{\Theta}.
\end{equation}
Here $ \mathcal{R} = (E B -a \Xi L_{z} )^{2} - R_{g} (\mathcal{K})$, $ \Theta = - (E A_{\pi/2} - \Xi L_{z})^{2} + \Theta_{g} \sin[\theta] ^{2} \mathcal{K} $ and $ \mathcal{K} = Q + (E A_{\pi/2} - \Xi L_{z})^{2}  $, with $ Q $ as the Carter's constant. \\
On writing $ \lambda = \frac{L_{z}}{E} $ and $ \eta = \frac{Q}{E^{2}} $ and then for solving $ \lambda $ and $ \eta $  by putting $ \mathcal{R} = 0 $ and $ \mathcal{R'} =0 $,
the celestial coordinates can be describe as, 
\begin{equation}\label{al}
	\alpha = -\frac{1}{Sin[\theta_0]} \dfrac{\Xi \left[\Xi \lambda - A_{\pi/2}\right]}{\sqrt{1-g^{2}\left[\eta + (\Xi \lambda -A_{\pi/2})\right]}}
\end{equation} 
\begin{equation}\label{be}
	\beta = \sqrt{\Theta_g} \dfrac{\sqrt{\eta-\frac{(\lambda Xi -A)^{2}}{\Theta_g Sin[\theta_0]^{2}} + (\lambda \Xi-A_{\pi/2})}}{\sqrt{1-g^{2}\left[\eta + (\Xi \lambda -A_{\pi/2})\right]}}.
\end{equation}

The equations \eqref{al} and \eqref{be} are the function of `r', $\mathcal{N}_{g}$, $\nu$, $e$ and $g$ involve in the spacetime  \eqref{metric} with mass $M=1$.  \\

\begin{figure*}[ht]
	\begin{center}
		
		\begin{tabular}{|c|c|c|}\hline
			\includegraphics[width= 5.5 cm, height= 4cm]{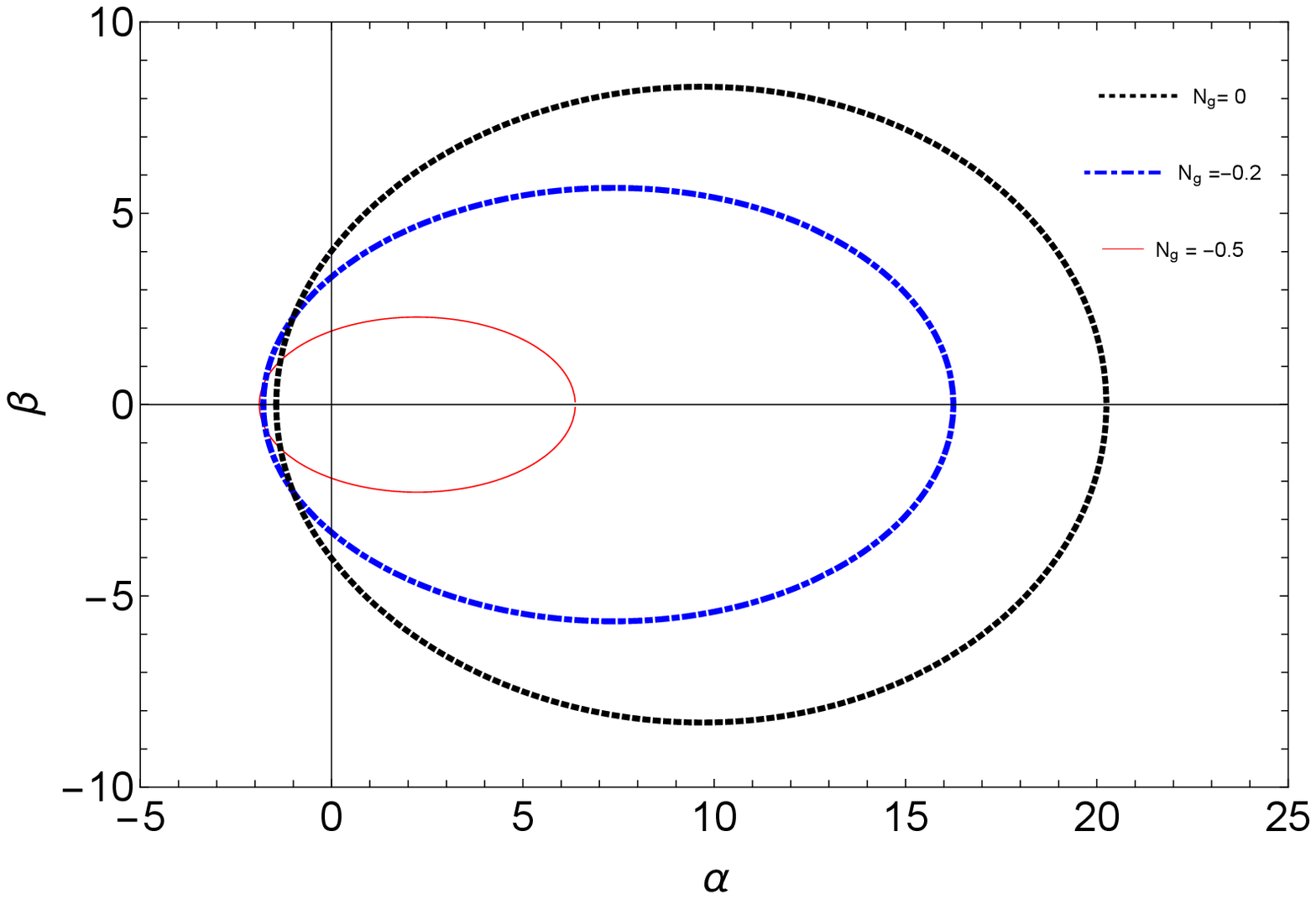}&
			\includegraphics[width= 5.5 cm, height= 4 cm]{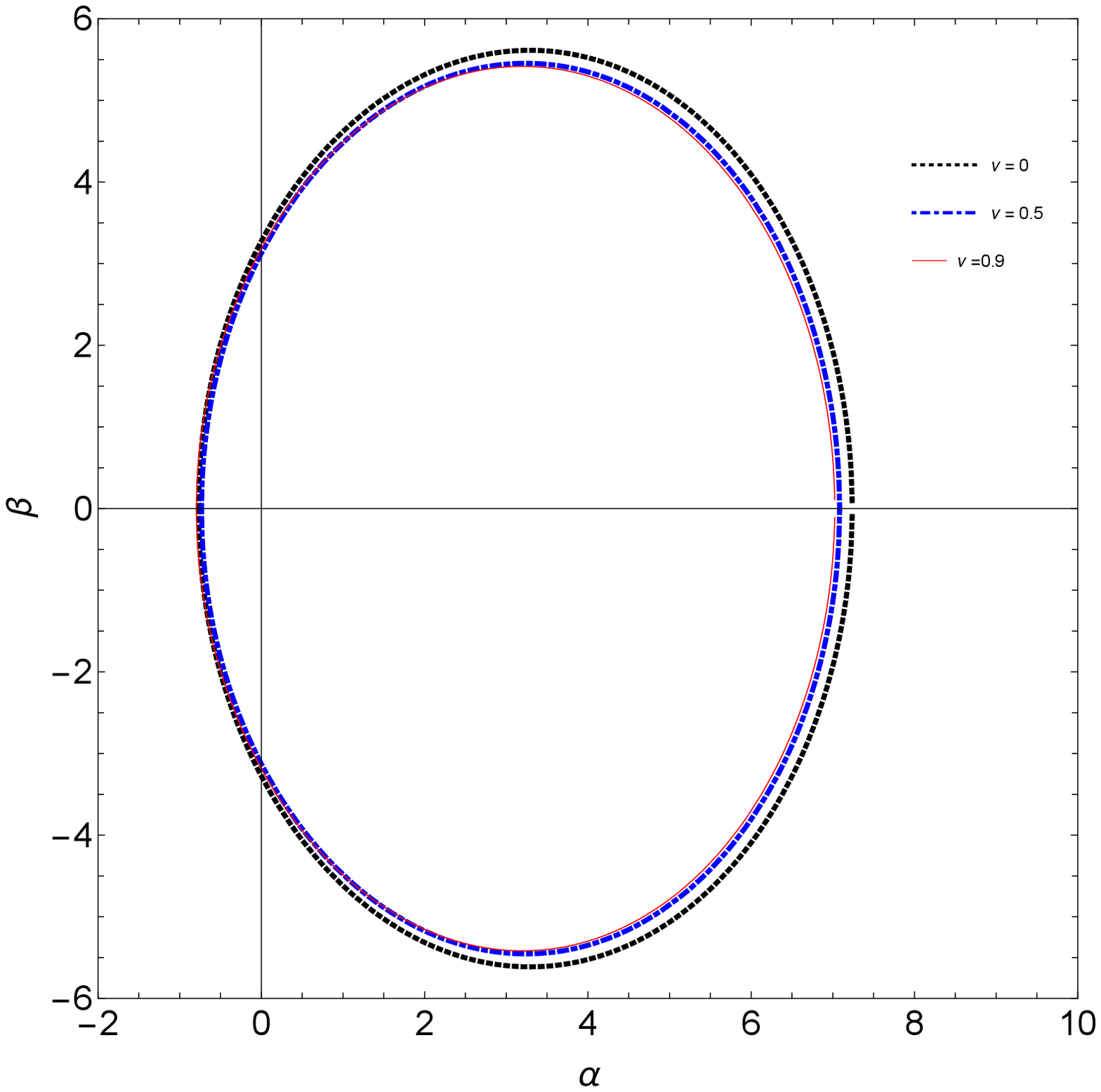}\\
			\hline
			\includegraphics[width= 5.5 cm, height= 4 cm]{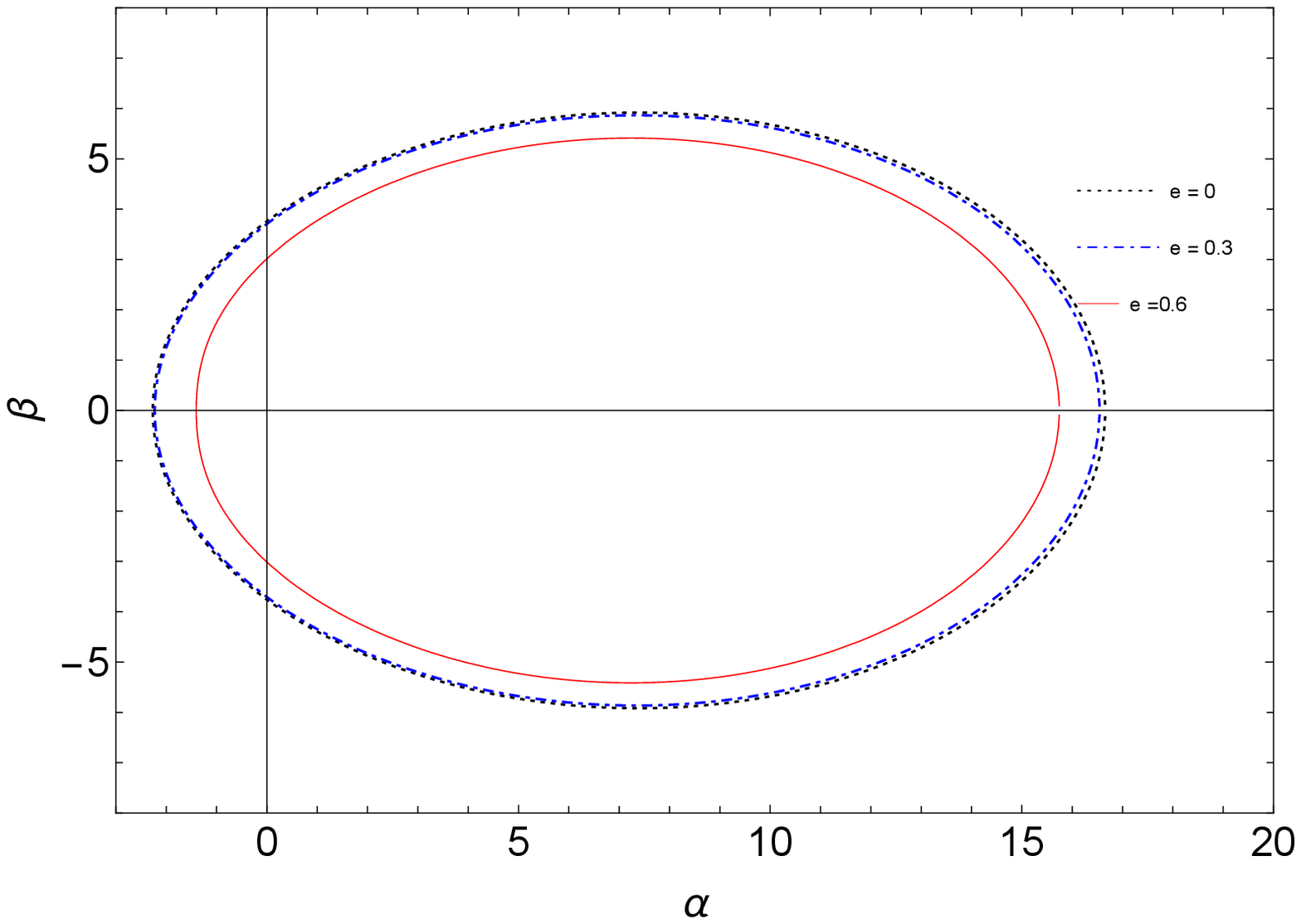}&
			\includegraphics[width= 5.5 cm, height= 4 cm]{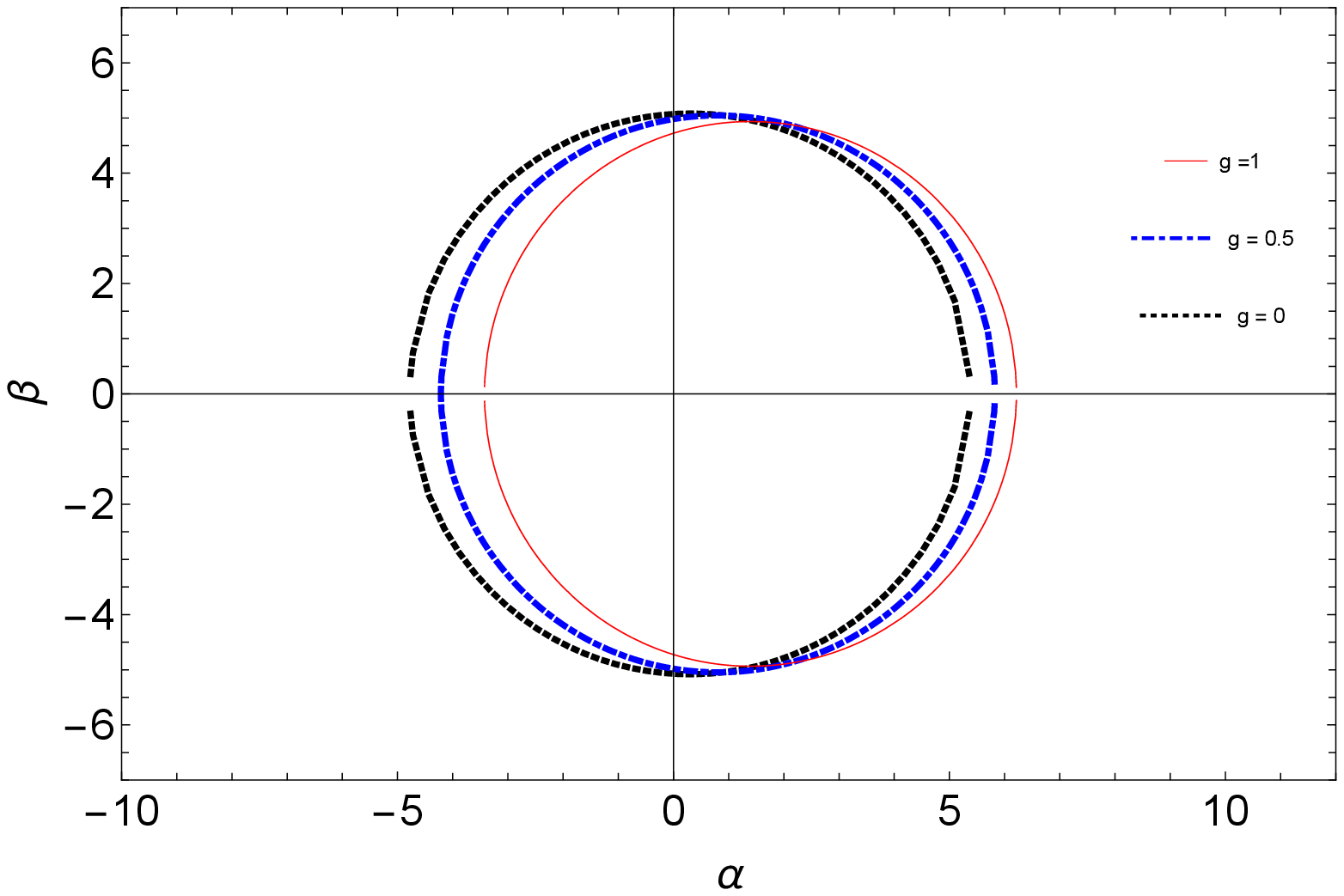}\\
			\hline
			
		\end{tabular} 
		\caption{\label{shadow all} The variation in photon rings for different values of   NUT charge $\mathcal{N}_{g}$ (Top left), magnetic charge $\nu$ (Top right), electric charge $e$ (Bottom left) and gauge coupling constant $g$ (Bottom right) $M=1$.}
	\end{center}
\end{figure*}

A pictorial view of photon ring is presented in  \figurename \ref{shadow all}. The $ \alpha $ and $ \beta $ is plotted for different values in top left corner in \figurename \ref{shadow all} where the photon ring is observed for different values of $ \mathcal{N}_{g} $. with, $ g = 1, a = 0.4, \nu = 0.1, e = 0.05 $ and $ M = 1 $. In top right corner the different values of magnetic charges are consider by putting other parameters constant $ g = 1, \mathcal{N}_{g} = 0.2, a = 0.4, e = 0.3 $ and $ M = 1 $. Bottom left corner represent photon ring with variation of $ e $ with  $ g = 1, \mathcal{N}_{g} = -0.2, \nu = 1.2, a = 0.4$  and $  M = 1 $. In bottom right corner the variation of g is studied when $ a = 0.1, \mathcal{N}_{g} = 0.1, \nu = 0.4, e = 0.2$ and $ M = 1 $. It is observed in present case of dyonic charge that the shadows are more oblate as compared to Schwarzschild BH and Kerr BH in general relativity (GR). 

\begin{figure*}[ht]
		
		\begin{tabular}{|c|c|c|}\hline
			\includegraphics[width= 5.5 cm, height=4 cm]{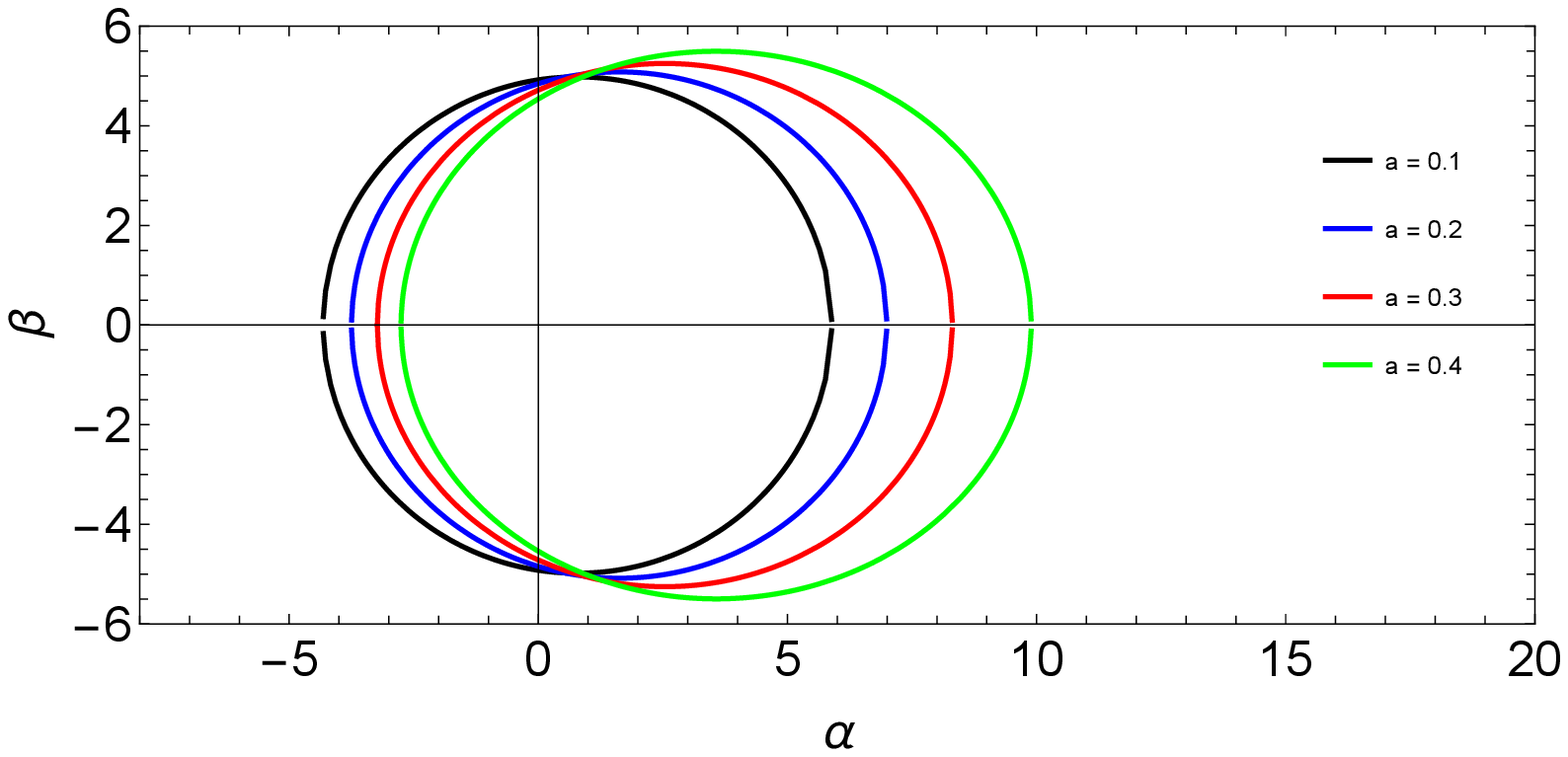}&
			\includegraphics[width= 5.5 cm, height=4 cm]{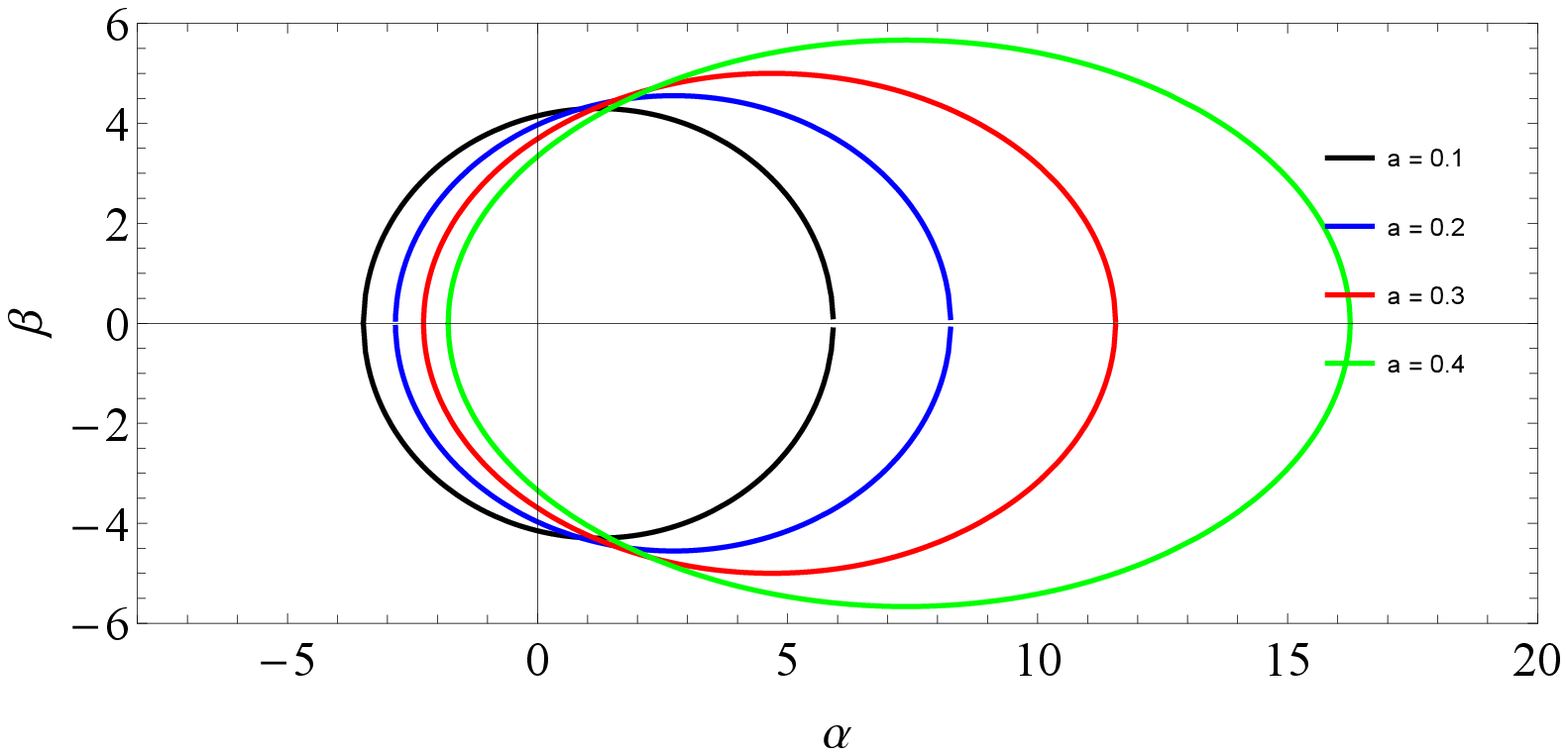}&
			\includegraphics[width= 5.5 cm, height=4 cm]{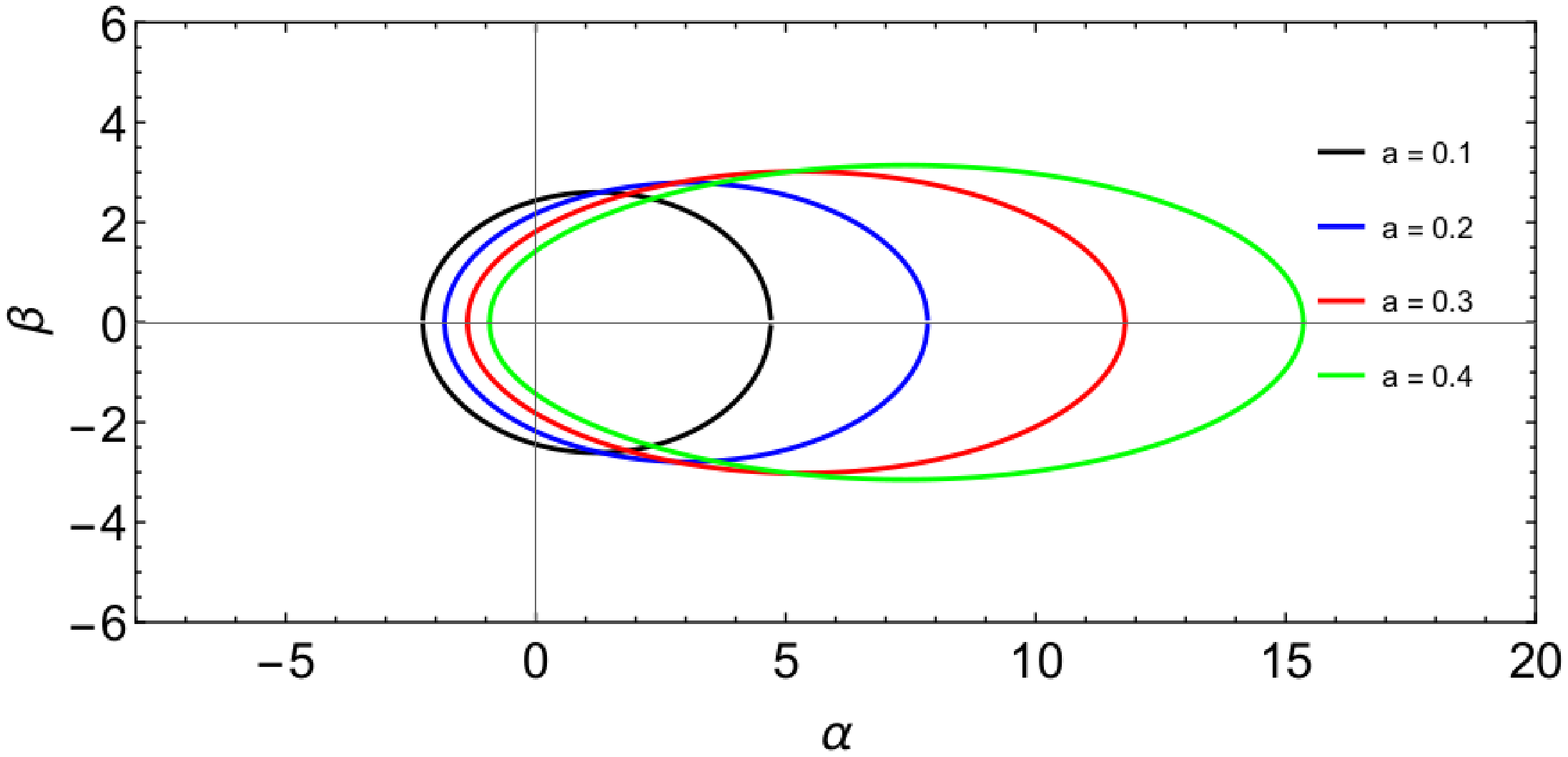}\\
			\hline
			\includegraphics[width= 5.5 cm, height=4 cm]{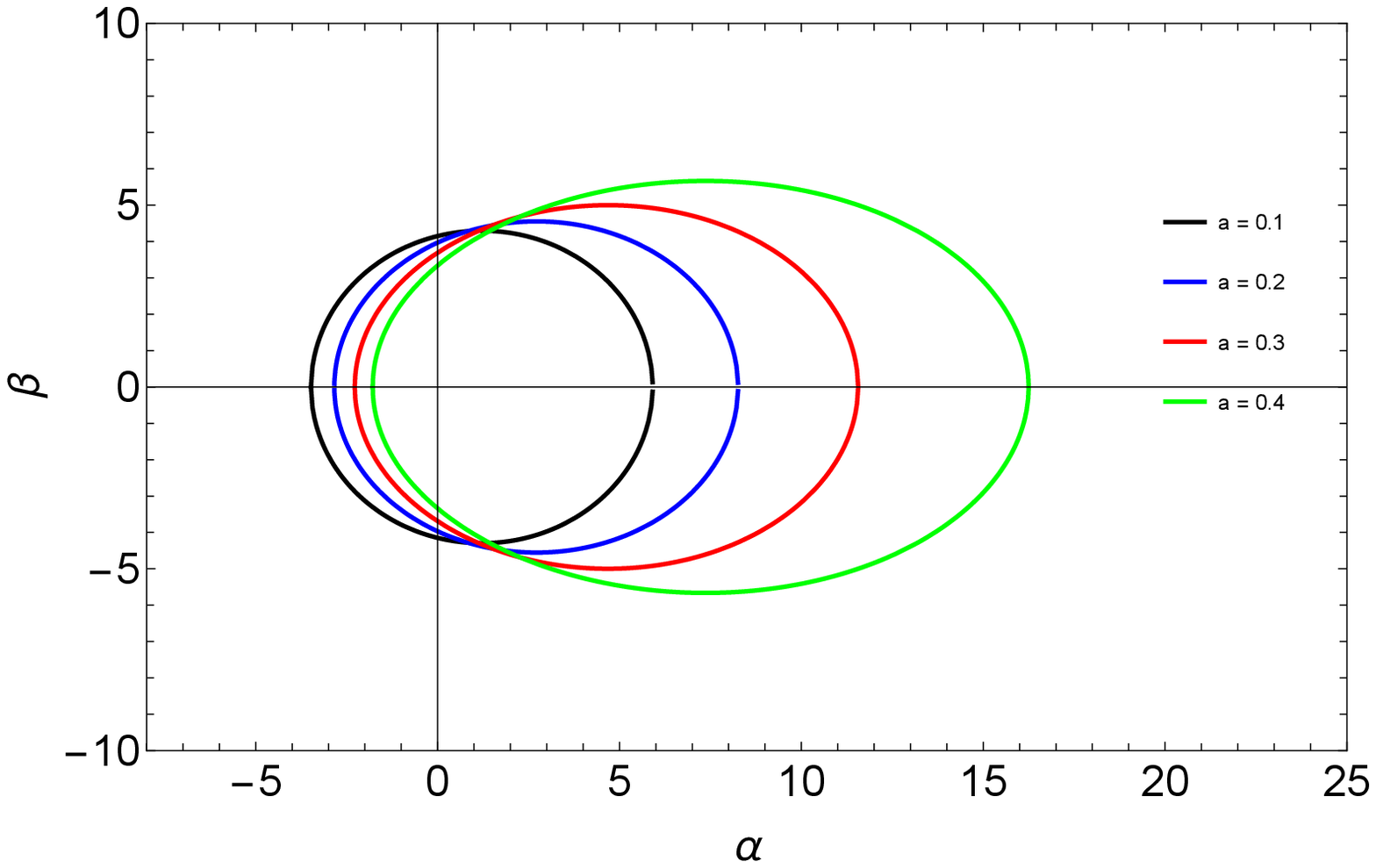}&
			\includegraphics[width= 5.5 cm, height=4 cm]{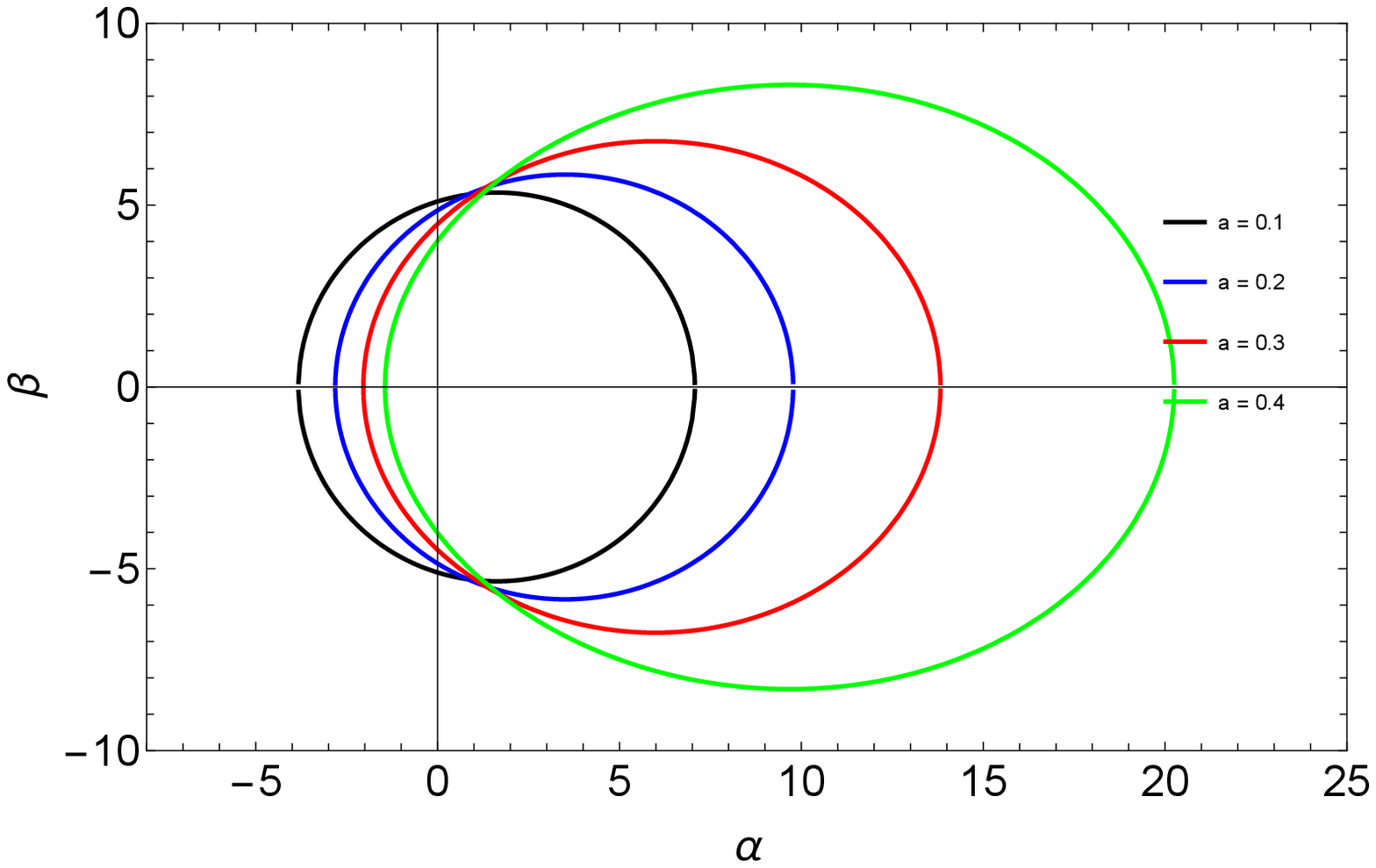}&
			\includegraphics[width= 5.5 cm, height=4 cm]{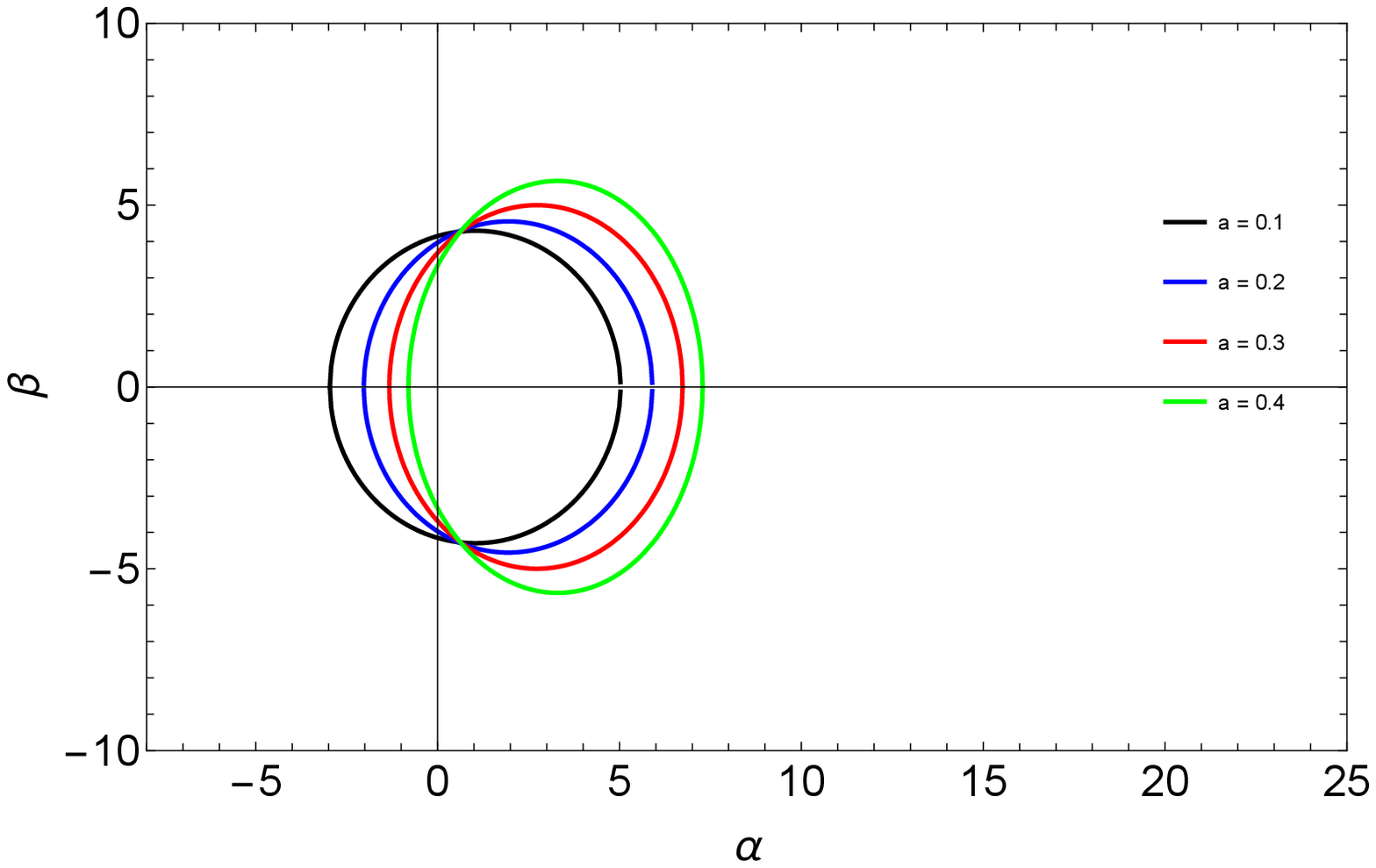}\\
			\hline
			\includegraphics[width= 5.5 cm, height=4 cm]{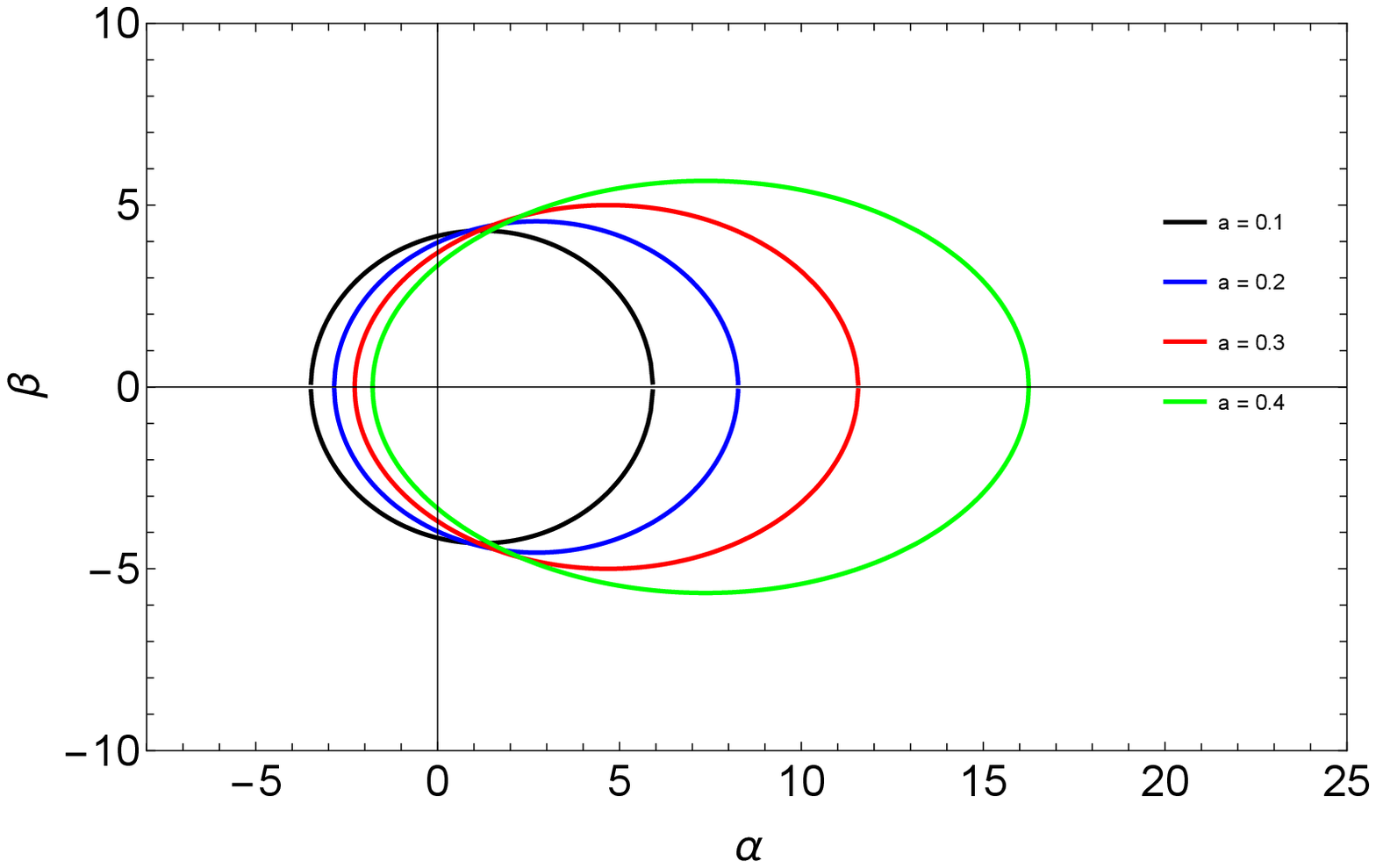}&
			\includegraphics[width= 5.5 cm, height=4 cm]{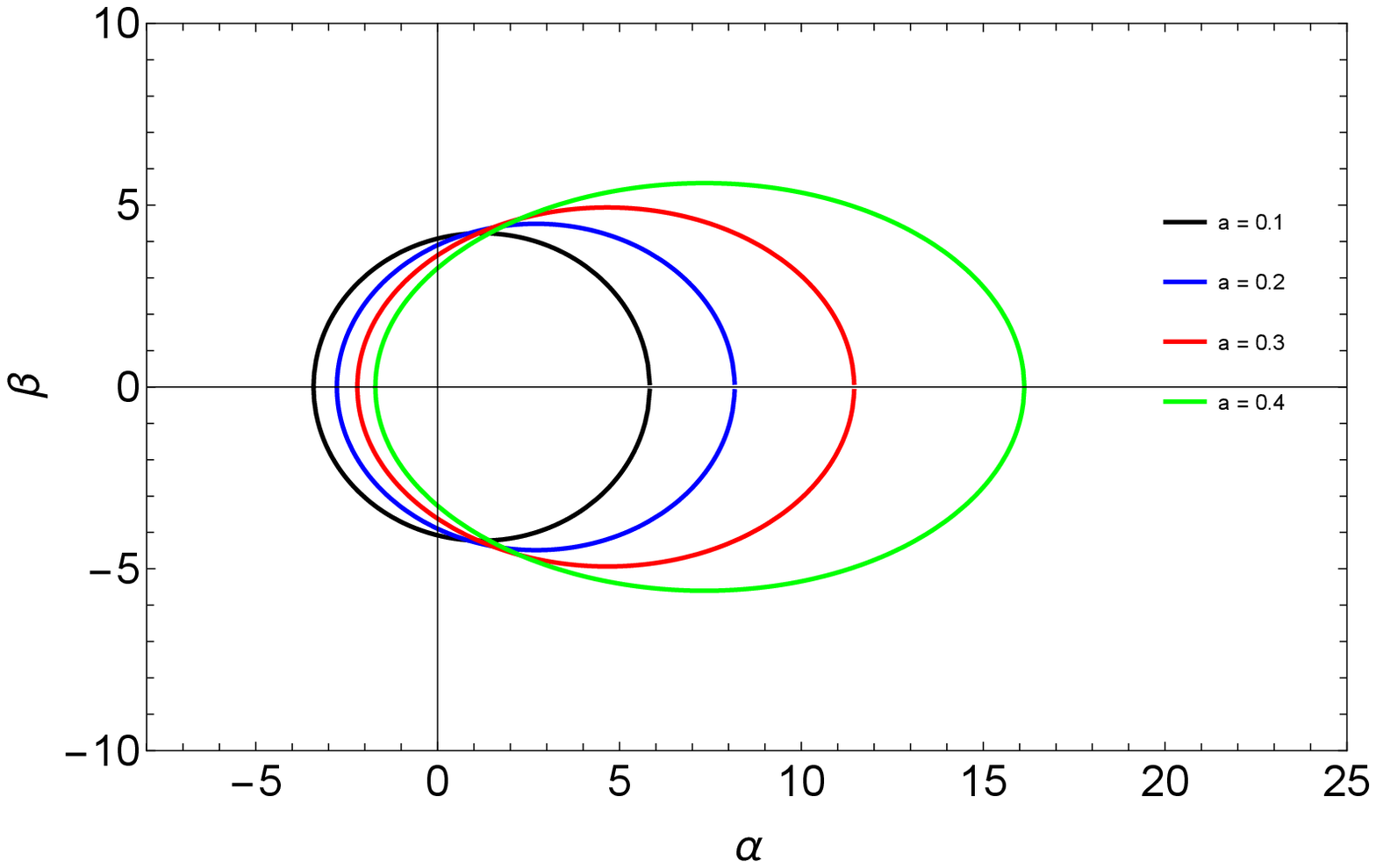}&
			\includegraphics[width= 5.5 cm, height=4 cm]{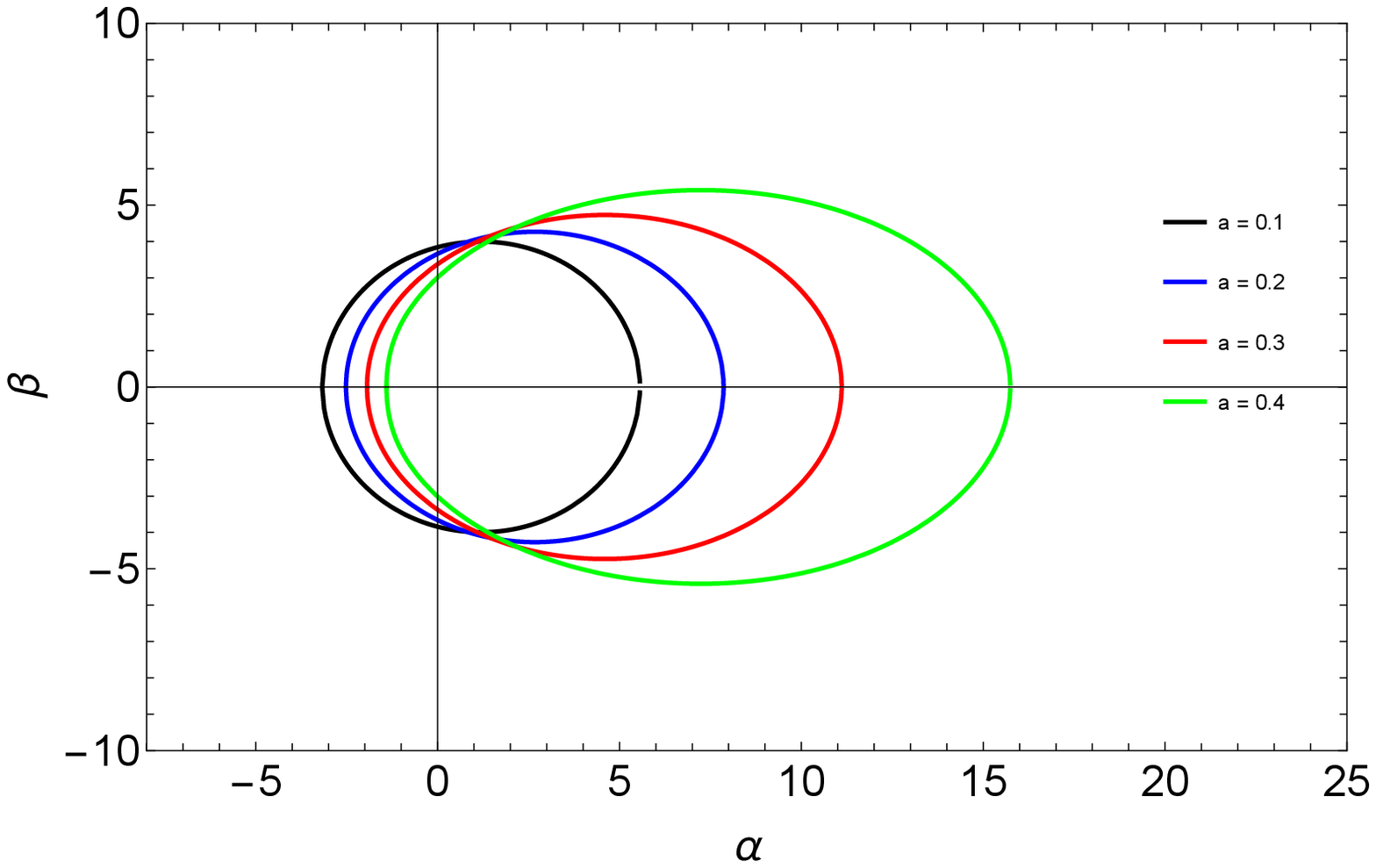}\\
			\hline
			\includegraphics[width= 5.5 cm, height=4 cm]{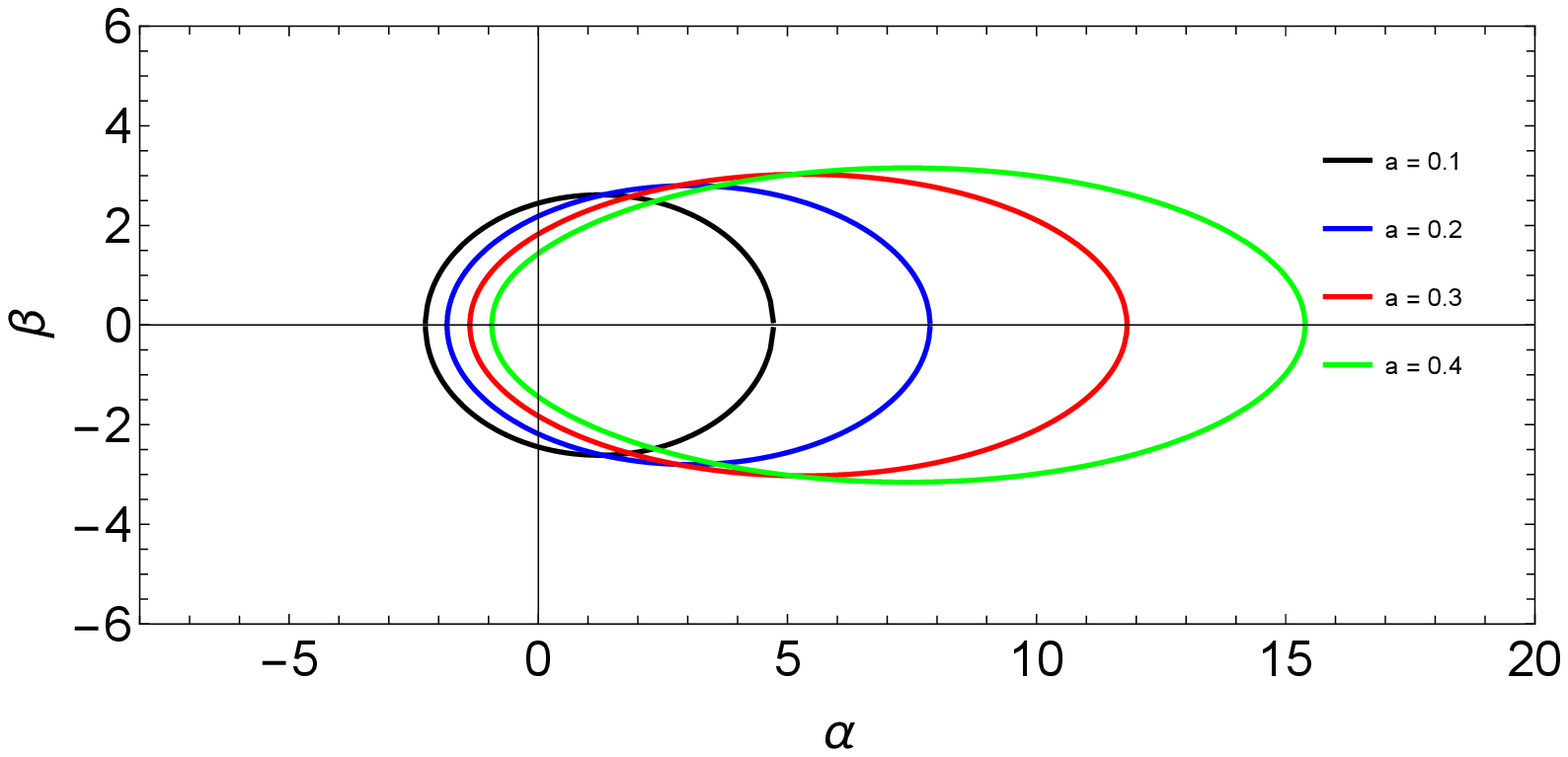}&
			\includegraphics[width= 5.5 cm, height=4 cm]{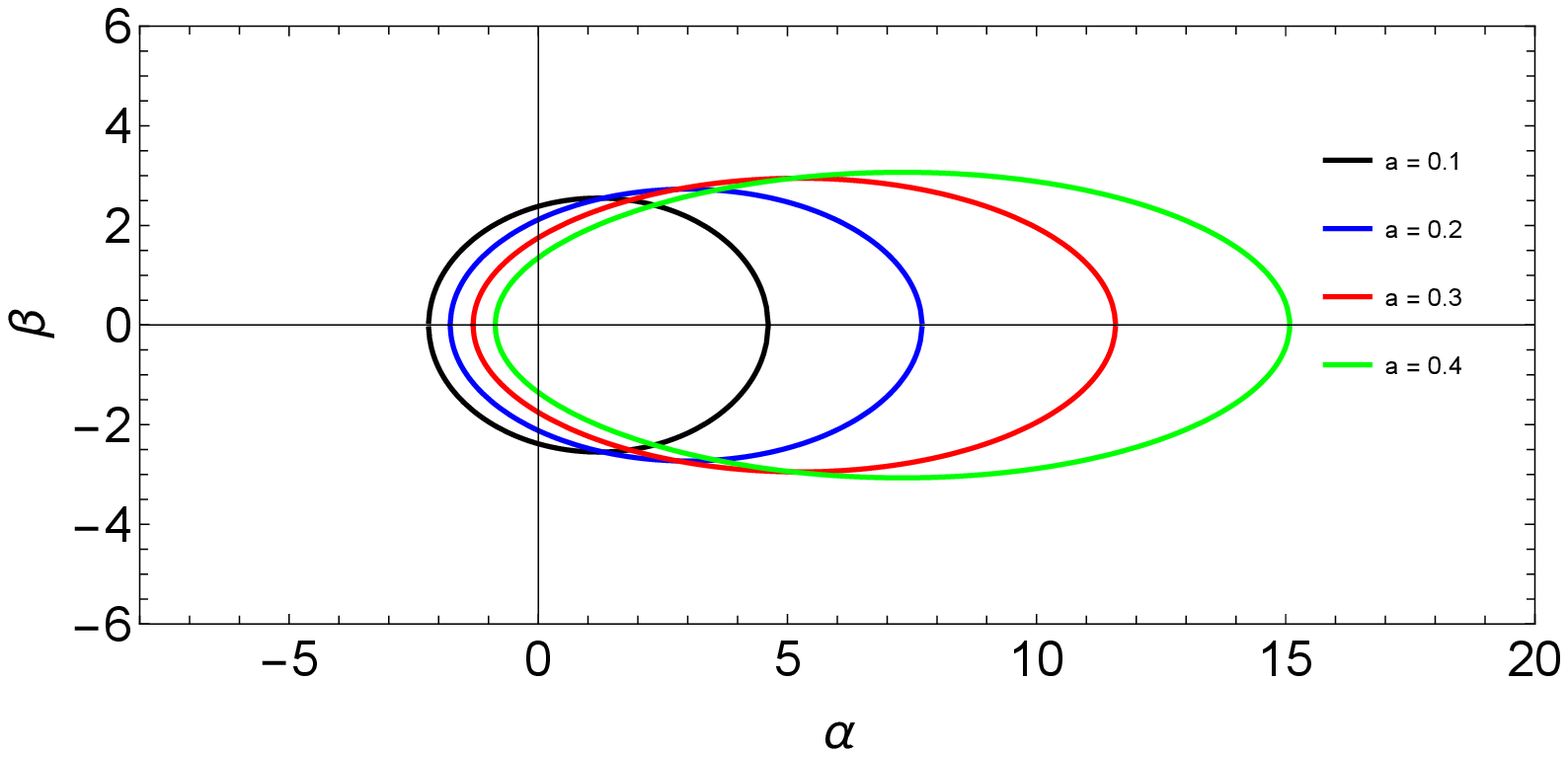}&
			\includegraphics[width= 5.5 cm, height=4 cm]{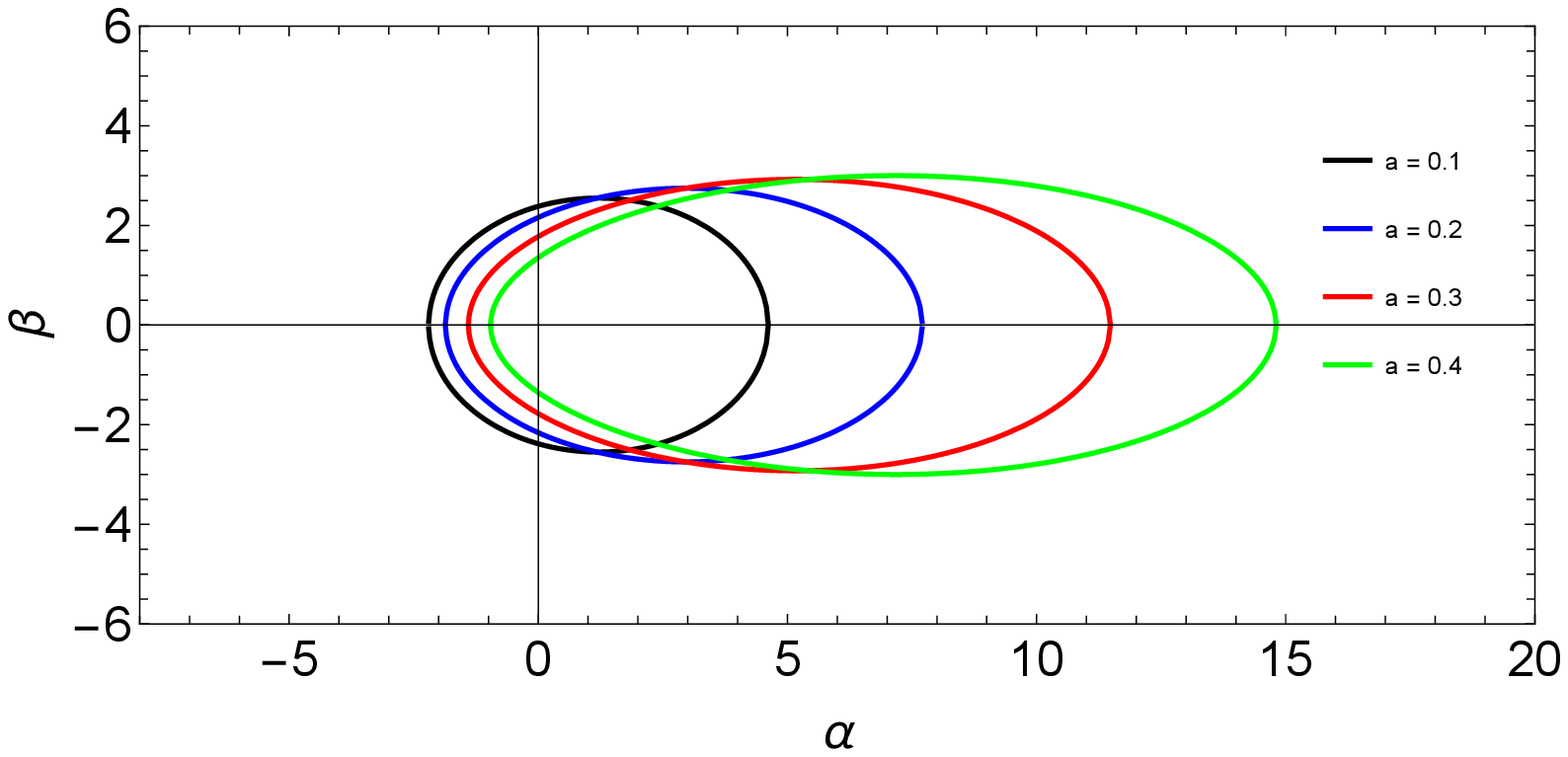}\\
			\hline
		\end{tabular} 
		\caption{\label{shadow a} The variation in photon rings for different values spin parameter `a' with different values of  gauge coupling constant $g$ (top panel), NUT charge $\mathcal{N}_{g} $ (second panel) , electric charge $e$  (third panel) and magnetic charge  $\nu $ (bottom panel) for  $M=1$.}
\end{figure*}

In \figurename\ref{shadow a}, the variation of photon ring with spin $ `a $' is analyzed with various parameter involve in. In uppermost panel of \figurename\ref{shadow a}, the coupling guage constant g is varied $ g =0.5 $ (left), $ g = 1 $ (middle) and  have $ g = 1 $ (right) while other parameters are $n = -0.2$, $\nu = 0.1$, $e = 0.05,$ and $M = 1$. In the second panel, the nut charge is varied  $ \mathcal{N}_{g} =-0.2 $ (left), $ \mathcal{N}_{g} = 0 $ (middle) and $ \mathcal{N}_{g} = 0.2 $ (right) the other parameters are $g = 1$, $\nu = 0.1$, $e = 0.05,$ and $M = 1$. In the third panel, electric charge is varied with $ e = 0 $ (left),  $ e = 0.3 $ (middle) and  $ e = 0.6 $ (right) the other parameters are $g = 1$, $\nu = 0.1$, $n = -0.2,$ and $M = 1$. In the 4 panel, the magnetic charge is varied with spin for $ \nu = 0 $ (left), $ \nu = 0.3 $ (middle) and  $ \nu = 0.6 $ (righ)t the other parameters are $g = 2$, $ e = 0.05$, $n = -0.2,$ and $M = 1$. 

In \figurename{\ref{shadow a}} and \figurename{\ref{shadow all}}, the effect of various BH parameter on photon ring is depicted. It can be observed that in supergravity regime, the shape of the shadow vary significantly as compared to BHs in GR.
\section{Thermodynamic structure}
In order to discuss the thermodynamic properties, we calculate various thermodynamic parameters below.
The mass of the BH is obtained by equating $ \Delta_{r_+} = 0$ as,  
\begin{widetext}
	\begin{eqnarray}
		M=\frac{a^2+\text{e1}^2-\text{Ng}^2+3 a^2 g^2 \text{Ng}^2-3 g^2 \text{Ng}^4+r^2+a^2 g^2 r^2+6 g^2 \text{Ng}^2 r^2+g^2 r^4-2 g^2 r^2 v^2}{2 r}.
	\end{eqnarray}
\end{widetext}

\begin{figure*}[ht]
	\begin{tabular}{|c|c|c|}\hline
		\includegraphics[width= 6 cm, height= 4 cm]{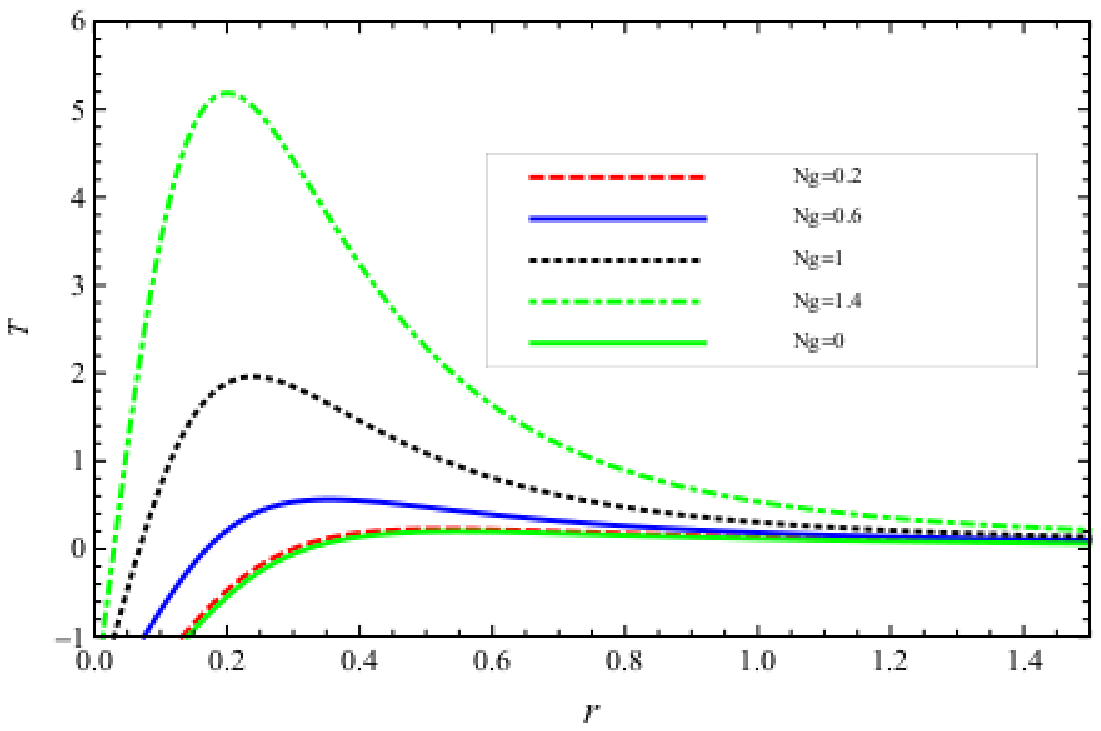}&
		\includegraphics[width= 6 cm, height= 4 cm]{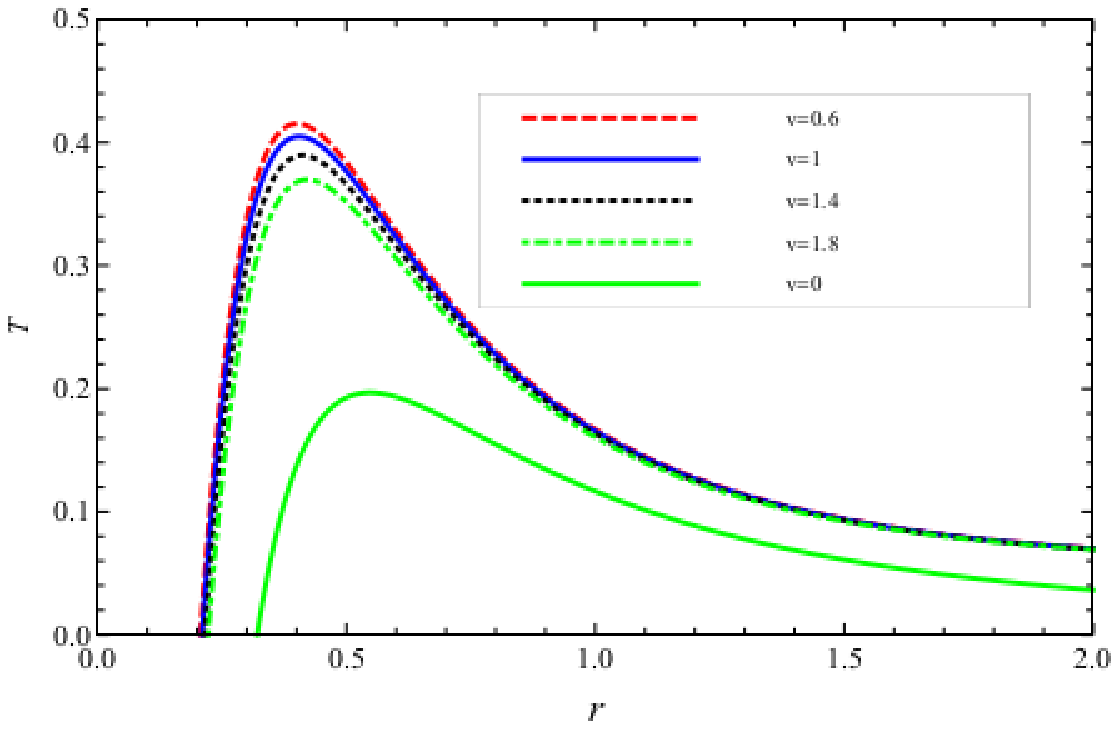}&
		
		\includegraphics[width= 6 cm, height= 4 cm]{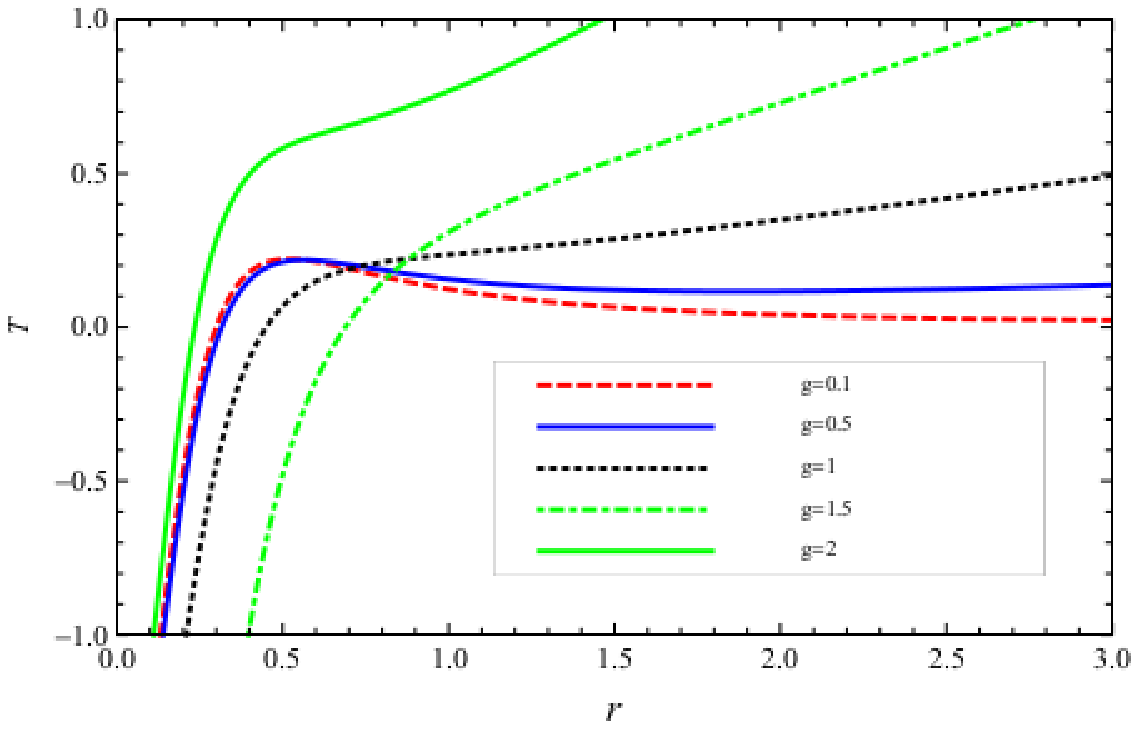}\\
		
		\hline
	\end{tabular} 
	\caption{\label{T} The variation of temperature with `r' for different values of i) NUT charge $\mathcal{N}_{g}$, ii) magnetic charge $\nu$ and iii) gauge coupling constant $g$ at $a=0.5, M=1$.}
	
\end{figure*}

The Hawking temperature associated with the BH is defined by $T=\kappa/2\pi$, where $\kappa$ is the surface gravity, 
\begin{equation}
	\kappa=\left\vert \frac{1}{2}\frac{R_g}{r^2+a^2}\right\vert_{(r_+)}.
\end{equation}
The Hawking temperature  using the surface gravity is then calculated as,
\begin{widetext}
\begin{eqnarray}
	T &=& \frac{1}{4\pi} \Big[\frac{-2m+2r+g^2(4r^3+2r(a^2+6N_g^2-2v^2))}{(a^2+r^2)} \nonumber \\ && - \frac{2r(a^2+e^2-N_g^2-2 m r+r^2+g^2(3N_g^2(a^2-N_g^2)+r^4+r^2(a^2+6N_g^2-2v^2)))}{(a^2+r^2)^2}\Big].
\end{eqnarray}

\end{widetext}

The temperature of the rotating dyonic BH in $\mathcal{N}_{g} = 2, U(1)^2$ gauged supergravity is plotted w. r. t. horizon radius in \figurename{\ref{T}} . It is observed that the Hawking temperature grows first to a maximum then decreases at constant value of gauge coupling constant. It turns out that the maximum value of the Hawking temperature increases with increasing NUT charge while decreasing with an increase the magnetic charge.  

Further an important and useful thermodynamic quantity associated with the BH horizon is its entropy. Using the Bekenstein-Hawking entropy relation, the entropy of the BH is defined using the area of event horizon $ S= A_H/ 4$. Therefore, to calculate the entropy, we obtain the area of the event horizon using the following definition\cite{poisson2004relativist},
\begin{widetext}
\begin{eqnarray}\label{AH}
	A_H &&= \int_{0}^{\pi} \int_{0}^{2\pi} \sqrt{g_{\theta \theta} g_{\phi \phi}} d\theta d\phi = 4 \pi  \sqrt{\left(-1+a g^2 (a+4 \text{Ng})\right)^2 \left((a+\text{Ng})^2+(r-v) (r+v)\right)^2}.
\end{eqnarray}
\end{widetext}

\begin{figure*}[ht]
	\begin{tabular}{|c|c|c|}\hline
		\includegraphics[width= 6 cm, height= 4 cm]{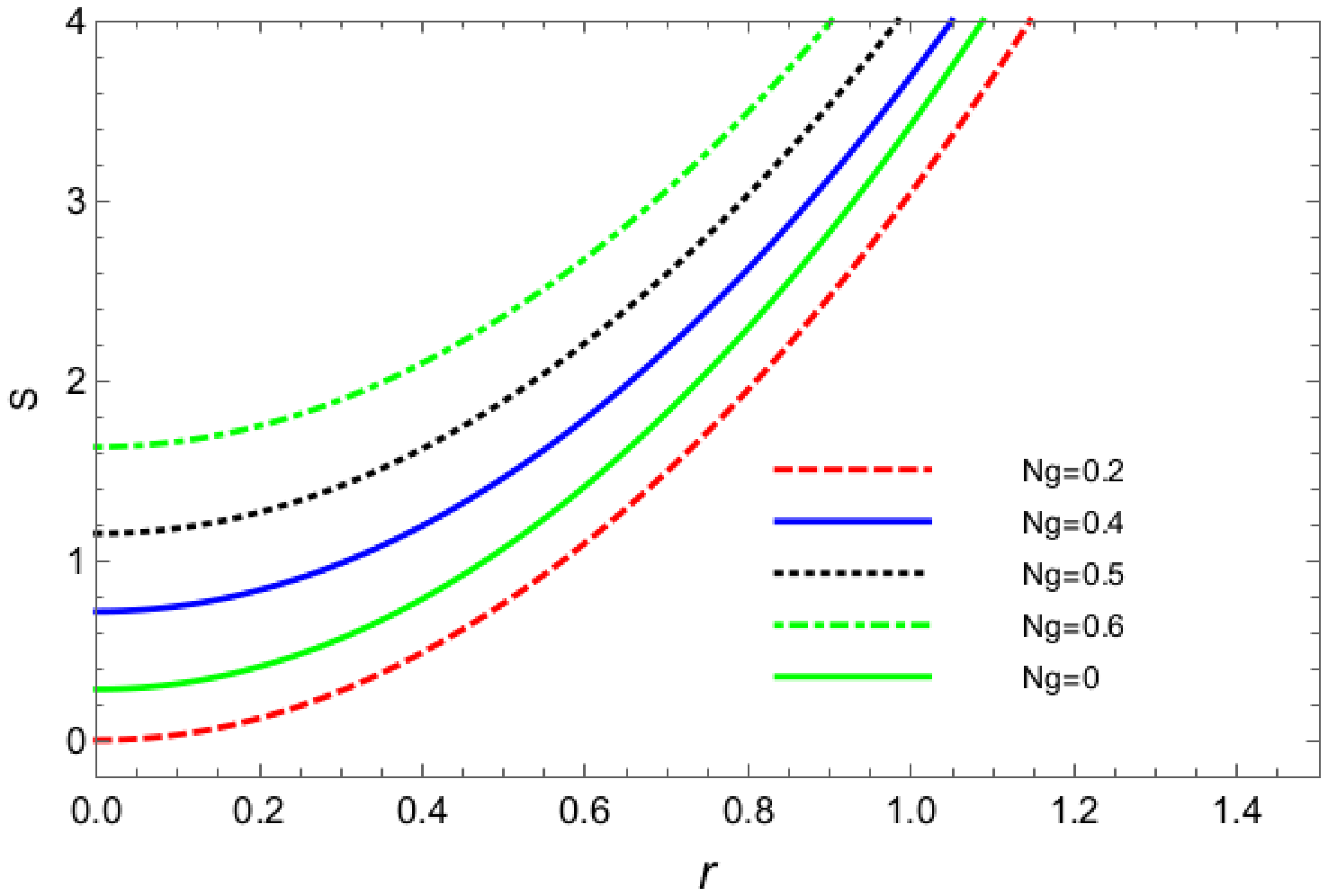}&
		\includegraphics[width= 6 cm, height= 4 cm]{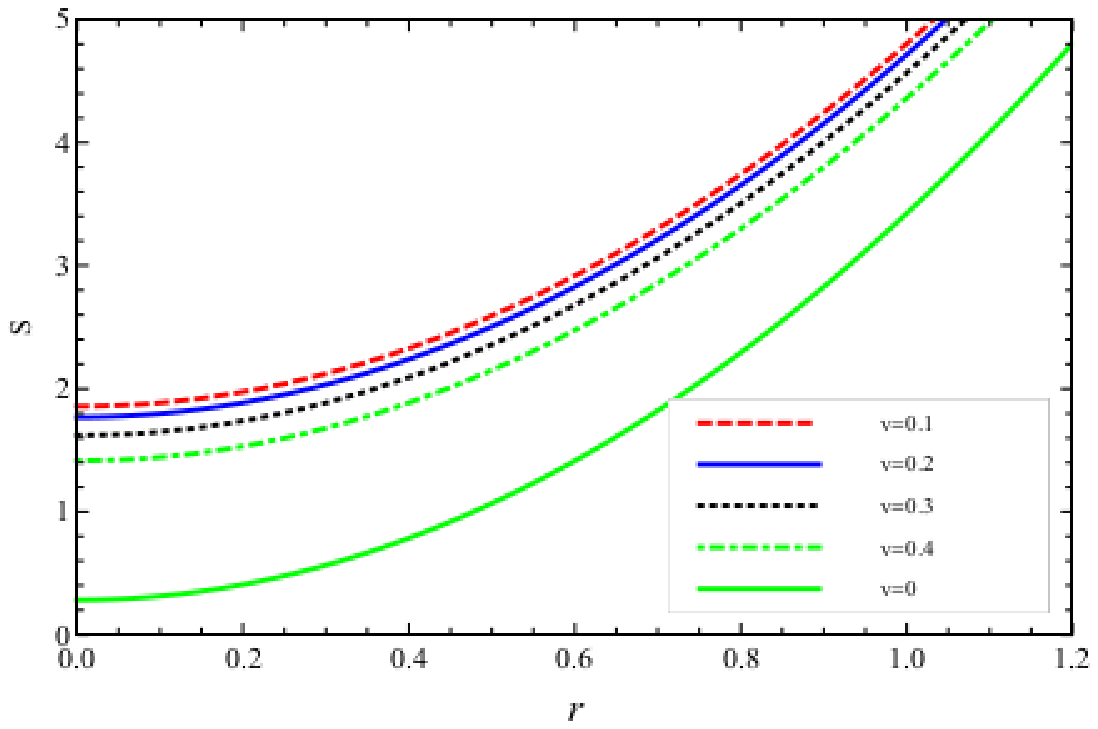}&
		
		\includegraphics[width= 6 cm, height= 4 cm]{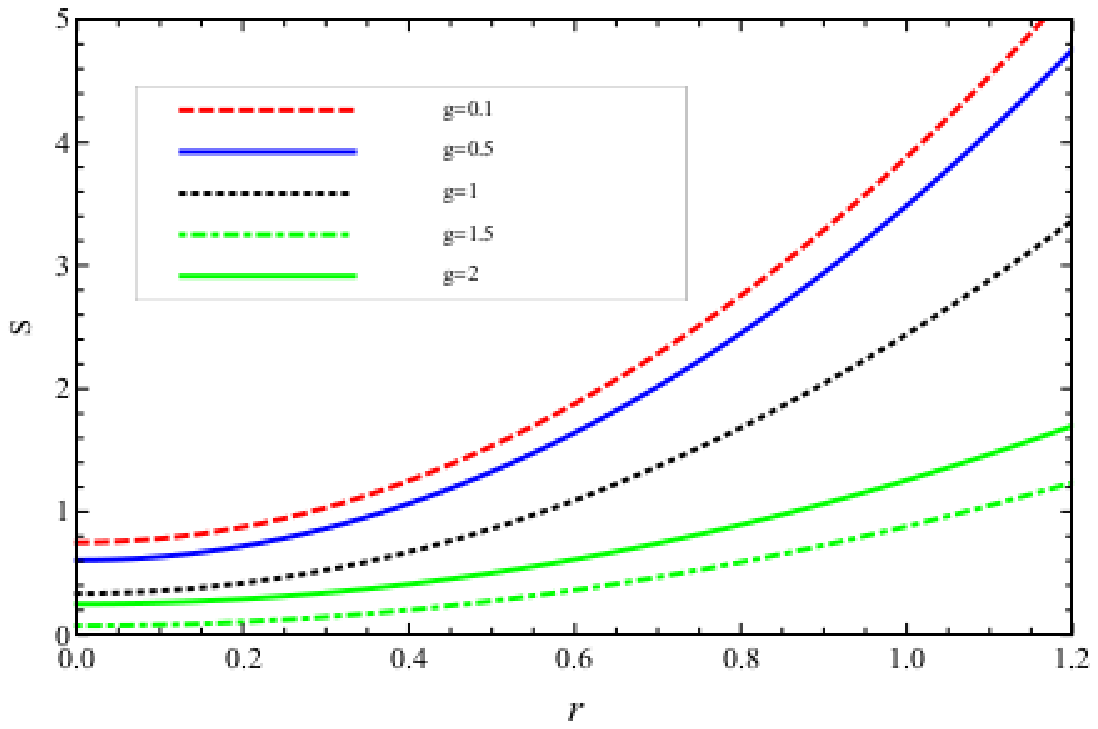}\\
		
		\hline
	\end{tabular} 
	\caption{\label{S}The variation of entropy with `r' for different values of i) NUT charge $\mathcal{N}_{g}$, ii) magnetic charge $\nu$ and iii) gauge coupling constant $g$ at $a=0.5, M=1$.}
\end{figure*}

The entropy for the BH spacetime \eqref{metric} is thus obtained as,\\
\begin{widetext}
\begin{eqnarray}
	S = \pi  \sqrt{\left(-1+a g^2 (a+4 \text{Ng})\right)^2 \left((a+\text{Ng})^2+(r-v) (r+v)\right)^2}.
\end{eqnarray}
\end{widetext}

The variation of entropy with horizon radius is plotted in \figurename{\ref{S}}, for different values of NUT charge $\mathcal{N}_{g}$, magnetic charge $\nu$ and gauge coupling constant $g$.

\begin{figure*}[ht]
	\begin{center}
		\begin{tabular}{|c|c|c|}\hline
			\includegraphics[width= 6 cm, height= 4 cm]{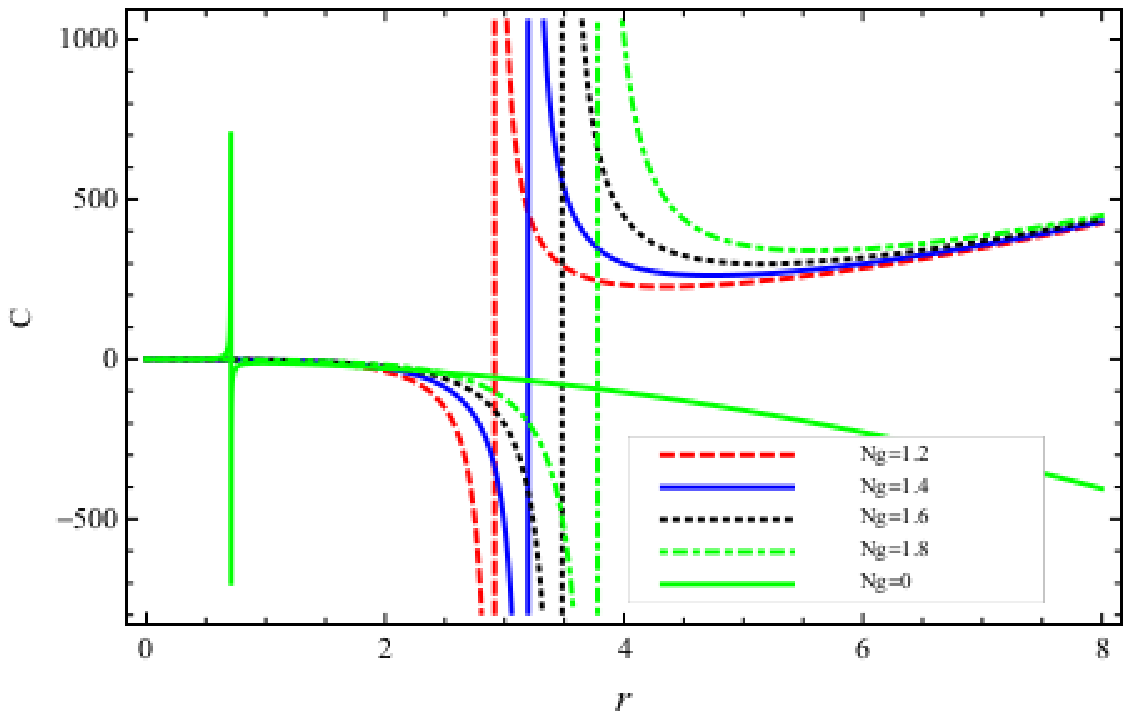}&
			\includegraphics[width= 6 cm, height= 4 cm]{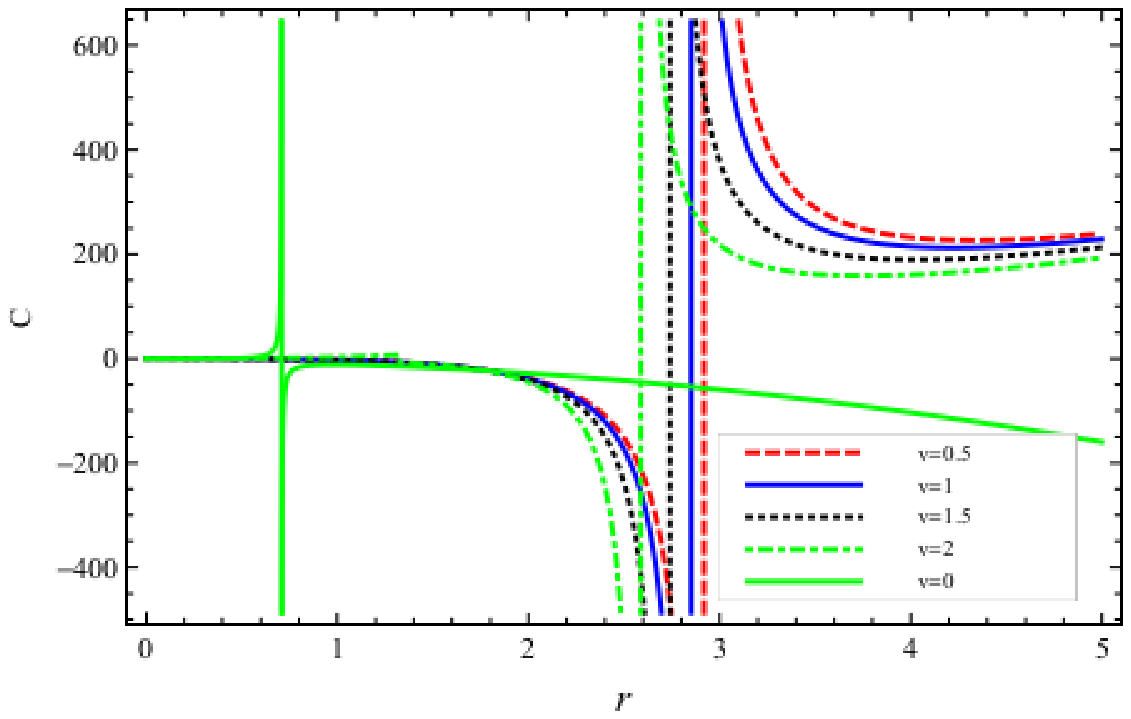}\\
			\hline    			
			\includegraphics[width= 6 cm, height= 4 cm]{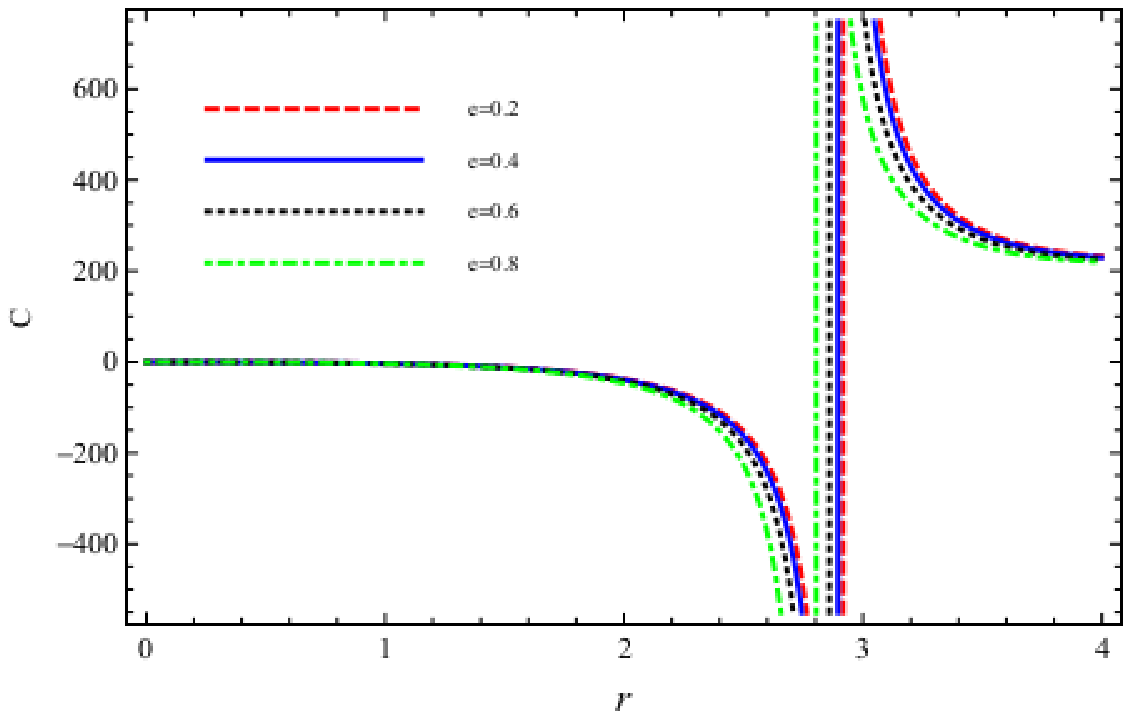}&
			\includegraphics[width= 6 cm, height= 4 cm]{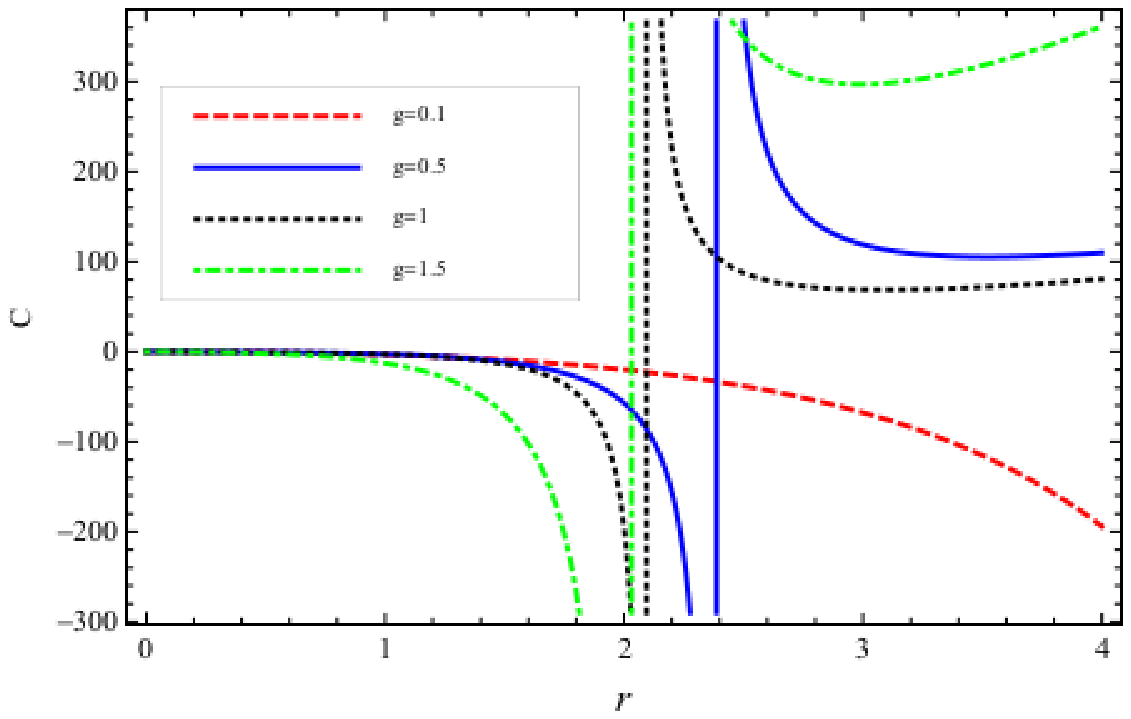}\\
			
			\hline
		\end{tabular} 
		\caption{\label{C}The variation of heat capacity with `r' for different values of   NUT charge $\mathcal{N}_{g}$ (Top left), magnetic charge $\nu$ (Top right), electric charge $e$ (Bottom left) and gauge coupling constant (g).}
	\end{center}
\end{figure*}

Next, to discuss the thermodynamic stability of the rotating dyonic BH in $\mathcal{N}_{g} = 2, U(1)^2$ gauged supergravity,  we compute the heat capacity of the BH. It is well known that the thermodynamic stability of the system is related to the sign of the heat capacity.  If the heat capacity is positive, then  the BH is stable; when it is negative, the BH is said to be unstable.  In a state of thermodynamic equilibrium, the BH can be considered as a system  that obeys the first law as below,
\begin{eqnarray}
	dM = TdS + \Omega dS + \phi d(e,N_g,v)
\end{eqnarray}
The heat capacity $C_{J,e,v,N_g}$ of the BH at constant $J$, $e$ , $v$, $N_g$ is calculated as,
  \begin{equation}
	C_{e,N_g,J} = \frac{\partial{M}}{\partial{T}}_{e,N_g,v,J}= \left(\frac{\partial{M}}{\partial{r_+}}\right)\left(\frac{\partial{r_+}}{\partial{T}}\right), \label{SH}
\end{equation}

the variation of heat capacity for different values of $\mathcal{N}_{g}$, $\nu$, $e$ and $g$ is presented in \figurename{\ref{C}}. 
\\
In this section, the thermodynamic quantities like the BH mass, Hawking temperature, entropy, specific heat are also calculated and their variation with BH parameters have being analyzed in detail. 
\section{Discussions, Summary/Conclusions  \& Future Directions}

We have studied the rotating dyonic BH in $\mathcal{N} = 2, U(1)^2$ gauged supergravity. In that direction, we have investigated
the phenomenon of BH shadows and various thermodynamic parameters with the variation of the BH parameters  $ N_{g}$, $ \nu $ , $ g $, $ e $ and $ a $. With this study  the following results can be drawn. 
\begin{itemize}
	\item It is observed as $ N_{g} $ becomes more negative the size of shadow increases when other parameters are kept constant. However, the variation of magnetic charge have trivial effect on the size of shadows. The coupling constant $ g $ changes the position of shadow as $ g $ increases from $ 0 $ to $ 1 $ the shadow shifts towards positive $ \alpha $, see bottom right box in \figurename{\ref{shadow all}}. When $ e $ is concerned it is observed that as the electric charge increased the size of the shadow which is oblate in shape decreases.
	\item It can be observed when the parameters in \ref{metric} are varied with a. In \figurename\ref{shadow a}, it is observed when $ a $ is varied with different values of g the shadow is oblate when $ g =0.5 $ and with increase in $ a $ the size of shadow increases. However, when we increase value of g the oblateness increases. When the different values of $ \mathcal{N}_{g} $ is considered it observed that at  $ \mathcal{N}_{g}  = -0.2 $ the shadow are oblate, also, as `$ a $' is increased the size of the photon ring increases. For, $  \mathcal{N}_{g} =0 $ the shape of shadow is more circular. However for $  \mathcal{N}_{g} = 0.2 $ the shape is prolate and the size increases with increase in $ a $. Moreover, when a is varied with different values of $ e $ and $ \nu $ there is no significant effect observed in shape and size of the shadow.
	\item The temperature ($ T $) and its variation with $ N_{g} $, $ \nu $ and $ g $ is studied. It is observed that the peak of maximum temperature increases as $ N_{g} $ increases. In case of magnetic charge, the peak of temperature decreases as $ \nu $ increases. However, the peak of maximum temperature decreases sharply when $ \nu $ is zero. The effect of $ g $ on $ T $ again show two types of characteristics as $ g $ increase from $ 0 $ to $ 1 $ the peak of $ T $ decreases after its maximum value it becomes constant. When $ g $ varies from $ 1.5 $ to $ 2 $ this variation shows opposite nature the value of $ T $ for $ g = 2 $ shows different nature, the value of $ T $ increases as $ r $ increases. For $ g = 1.5 $ the value of $ T $ increases as $ r $ increase but the value of $ T $ is minimum with respect to last values of $ g $ when $ r $ lies between $ 0 $ to $ 0.6 $ after it increase with $ r $.
	\item The entropy ($ S $) and its variation with $ N_{g} $, $ \nu $, and $ g $ is studied. It can be observed that as $ N_{g} $ increases the value of entropy shifted towards the smaller values of $ r $. Similar trend can be observed with the $ \nu $ as $ \nu $ increases the value of entropy shifted toward smaller values of $ r $. However, the effect of coupling constant on entropy show two types of characteristics as $ g $ increase from $ 0 $ to $ 1 $ the entropy decreases for particular value of $ r $. When $ g $ varies from $ 1.5 $ to $ 2 $ this variation shows opposite nature the value of entropy increases for a particular value of $ r $ as $ g $ changes from $ 1.5 $ to $ 2 $.
	\item The thermodynamic stability of black holes condition is based on the sign of heat capacity. The change of sign could happen whether when heat capacity meets root or divergency. The heat capacity divergencies is considered as phase transition points. The negativity of the heat capacity represents unstable solutions, whereas its positive value is related to stable state. Heat capacity as a function of horizon radius for different values of parameter  $ N_{g} $, $ \nu $, $ e $ and $ g $ shows that for all the parameters there exist a point where there is a phase transition from the stable to unstable state. It can also be seen that for the increase in NUT charge from $ N_{g} = 1.2 $ to $ N_{g} = 1.8 $ the point of transition shifted towards the larger value of $ r $. The effect of magnetic charge ($ \nu $) on transition point is as $ \nu $ increase from $ \nu = 0.5 $ to $ \nu = 2 $ the transition shifted toward smaller values of $ r $. Similarly the effect of electric charge ($ e $) shows as $ e $ increase the value of transition point shifted toward smaller value of $ r $. The variation of transition point with coupling constant is interesting because when $ g $ changes from $ 0.1 $ to $ 0.5 $ the $ C $ significantly fall toward smaller value of $ r $ however, as $ g $ vary from $ 0.5 $ to $ 1.5 $ the change in transition point with $ r $ is smaller as compared with earlier values but the characteristic of variation remain same as $ g $ increases the $ C $ fall at smaller value of $ r $.

\end{itemize}
All the above mentioned results obtained reduce to the case of usual Kerr-Newman BH, Kerr BH and Schwarzs-child BH in GR  in the prescribed limit. Further, in near future the study of shadows and gravitational lensing will presented for the Rotating Dyonic Black Hole in $\mathcal{N}_{g} = 2, U(1)^2$ gauged supergravity in the presence of plasma background and the results will be then compared with Schwarzschild BH and Kerr BH in GR.

\begin{acknowledgements}
	HN and PS would like to thank the Science and Engineering Research Board (SERB), DST, New Delhi for financial support through the grant number EMR $ /2017 /000339$. Author, UP, would like to thank University Grant Commission (UGC), New Delhi for DSKPDF through grant $No. F.4-2/2006(BSR)/PH/18-19/0009$. The authors 
	acknowledge the facilities used at IUCAA Centre for Astronomy Research and Development (ICARD), Gurukula Kangri (Deemed to be University), Haridwar, India. The authors are also thankful to Dr. Rajibul Shaikh and Anik Rudra for useful discussions during an early stage of this work.
\end{acknowledgements}
\bibliographystyle{unsrt}
\bibliography{dyonicbib}

\end{document}